\PassOptionsToPackage{dvipsnames,table}{xcolor}

\documentclass[nonacm,acmsmall]{acmart}

\AtBeginDocument{%
  \providecommand\BibTeX{{%
    Bib\TeX}}}

\setcopyright{acmcopyright}
\copyrightyear{2018}
\acmYear{2018}
\acmDOI{XXXXXXX.XXXXXXX}

\acmConference[Conference acronym 'XX]{Make sure to enter the correct
  conference title from your rights confirmation emai}{June 03--05,
  2018}{Woodstock, NY}

\acmPrice{15.00}
\acmISBN{978-1-4503-XXXX-X/18/06}

\AtBeginDocument{%
  \providecommand\BibTeX{{%
    \normalfont B\kern-0.5em{\scshape i\kern-0.25em b}\kern-0.8em\TeX}}}

\setcopyright{acmcopyright}
\copyrightyear{2024}
\acmYear{2024}
\acmDOI{XXXXXXX.XXXXXXX}

\acmPrice{15.00}
\acmISBN{978-1-4503-XXXX-X/18/06}

\usepackage{graphicx} % Required for inserting images
\usepackage{xspace}
\usepackage{listings}
\usepackage[skins]{tcolorbox}
\usepackage{algorithm2e}
\usepackage{tabularx}
\usepackage{soul}
\usepackage{multirow}
\usepackage{pifont}
\usepackage{arydshln}

\colorlet{LightCyan}{Cyan!60}
\colorlet{LightOrchid}{Orchid!60}

\newcommand{\ttt}{\ensuremath{\mathsf{True}}}

\newcommand{\mcomment}[1]{{}}
\newcommand{\myparagraph}[1]{\vspace*{2mm}\noindent\textit{\textbf{#1}.}}

\newcommand{\cyanparagraph}[1]{\vspace*{2mm}\noindent\hlcyaninline{#1}}
\newcommand{\orchidparagraph}[1]{\vspace*{2mm}\noindent\hlorchidinline{#1}}

\newcommand{\sysname}{\ensuremath{\mathsf{CodePlan}}\xspace}

\newcommand{\zug}[1]{\ensuremath{\langle {#1} \rangle}}

\newcommand{\todo}[1]{{\color{red} #1}}
\newcommand{\code}[1]{{\texttt{#1}}}

\newcommand{\Dprime}{$\text{D}^\prime$\xspace}

\newcommand{\mylstinline}[1]{\lstinline[basicstyle=\fontfamily{pcr}\selectfont\small]{#1}}

\newcommand{\hlcyaninline}[1]{\sethlcolor{LightCyan}{\fontfamily{pcr}\selectfont\small\hl{#1}}}

\newcommand{\hlorchidinline}[1]{\sethlcolor{LightOrchid}{\fontfamily{pcr}\selectfont\small\hl{#1}}}

\definecolor{darkgreen}{rgb}{0, 0.5, 0}

\newcommand{\prompt}[2]{%
    \begin{tcolorbox}[
        space to upper,
        colframe=black!80!white,
        colback=white,
        halign=left,
        valign=top,
        halign lower=flush left,
        title=#1
    ]
    #2
    \end{tcolorbox}%
}

\newcommand*\circled[1]{\tikz[baseline=(char.base)]{
            \node[shape=circle,fill,inner sep=0.5pt] (char) {\textcolor{white}{#1}};}}

\lstdefinestyle{diff}{
    morecomment=[l][\color{Green}]{+\ },
    morecomment=[l][\color{Red}]{-\ }
}

\lstalias{no_lang}{}

\DeclareRobustCommand{\hlcyan}[1]{{\sethlcolor{LightCyan}\hl{#1}}}
\DeclareRobustCommand{\hlorchid}[1]{{\sethlcolor{LightOrchid}\hl{#1}}}

\title{\sysname: Repository-level Coding using LLMs and Planning}

\author{Ramakrishna Bairi}
\email{rbairi@microsoft.com}
\affiliation{%
  \institution{Microsoft Research}
  \country{India}
}

\author{Atharv Sonwane}
\email{t-asonwane@microsoft.com}
\affiliation{%
  \institution{Microsoft Research}
  \country{India}
}

\author{Aditya Kanade}
\email{kanadeaditya@microsoft.com}
\affiliation{%
  \institution{Microsoft Research}
  \country{India}
}

\author{Vageesh D C}
\email{vachand@microsoft.com}
\affiliation{%
  \institution{Microsoft Research}
  \country{India}
}

\author{Arun Iyer}
\email{ariy@microsoft.com}
\affiliation{%
  \institution{Microsoft Research}
  \country{India}
}

\author{Suresh Parthasarathy}
\email{supartha@microsoft.com}
\affiliation{%
  \institution{Microsoft Research}
  \country{India}
}

\author{Sriram Rajamani}
\email{sriram@microsoft.com}
\affiliation{%
  \institution{Microsoft Research}
  \country{India}
}

\author{B. Ashok}
\email{bash@microsoft.com}
\affiliation{%
  \institution{Microsoft Research}
  \country{India}
}

\author{Shashank Shet}
\email{t-sshet@microsoft.com}
\affiliation{%
  \institution{Microsoft Research}
  \country{India}
}

\begin{document}

\begin{abstract}

Software engineering activities such as package migration, fixing errors reports from static analysis or testing, and adding type annotations or other specifications to a codebase, involve pervasively editing the entire repository of code. We formulate these activities as {\em repository-level coding} tasks.

Recent tools like GitHub Copilot, which are powered by Large Language Models (LLMs), have succeeded in offering high-quality solutions to localized coding problems. Repository-level coding tasks are more involved and cannot be solved directly using LLMs, since code within a repository is inter-dependent and the entire repository may be too large to fit into the prompt.
We frame repository-level coding as a planning problem and present a task-agnostic framework, called \sysname\ to solve it.
\sysname\ synthesizes a multi-step \emph{chain of edits} (plan), where each step results in a call to an LLM on a code location with context
derived from the entire repository, previous code changes and task-specific instructions.
\sysname\ is based on a novel combination of an incremental dependency analysis, a change may-impact analysis and an adaptive planning algorithm.

We evaluate the effectiveness of \sysname\ on two repository-level tasks: package migration (C\#) and temporal code edits (Python). Each task is evaluated on multiple code repositories, each of which requires inter-dependent changes to many files (between 2--97 files). Coding tasks of this level of complexity have not been automated using LLMs before. Our results show that \sysname has better match with the ground truth compared to baselines. \sysname is able to get 5/6 repositories to pass the validity checks (e.g., to build without errors and make correct code edits) whereas the baselines (without planning but with the same type of contextual information as \sysname) cannot get any of the repositories to pass them.
We will release our data and evaluation scripts at \url{https://aka.ms/CodePlan}.
\end{abstract}




\maketitle

\section{Introduction}

\begin{figure}
\begin{minipage}{0.53\linewidth}

\raggedright
We use a Complex Numbers library that had the following edit -
\begin{lstlisting}[style=diff,frame=single]
+ class Complex {
+   float real;
+   float imag;
+   dict<string, string> metadata;
+ }
    
- tuple<float, float> create_complex(float a, float b) 
+ Complex create_complex(float a, float b, dict metadata) 
\end{lstlisting}
Modify the code repository in accordance with this change.   
\vspace*{-7pt}
\caption{Task instruction to migrate a code repository due to an API change in the Complex Numbers library.}
\label{fig:complexnumber}
\end{minipage}
\hfill
\begin{minipage}{0.44\linewidth}
\centering
\includegraphics[width=\linewidth]{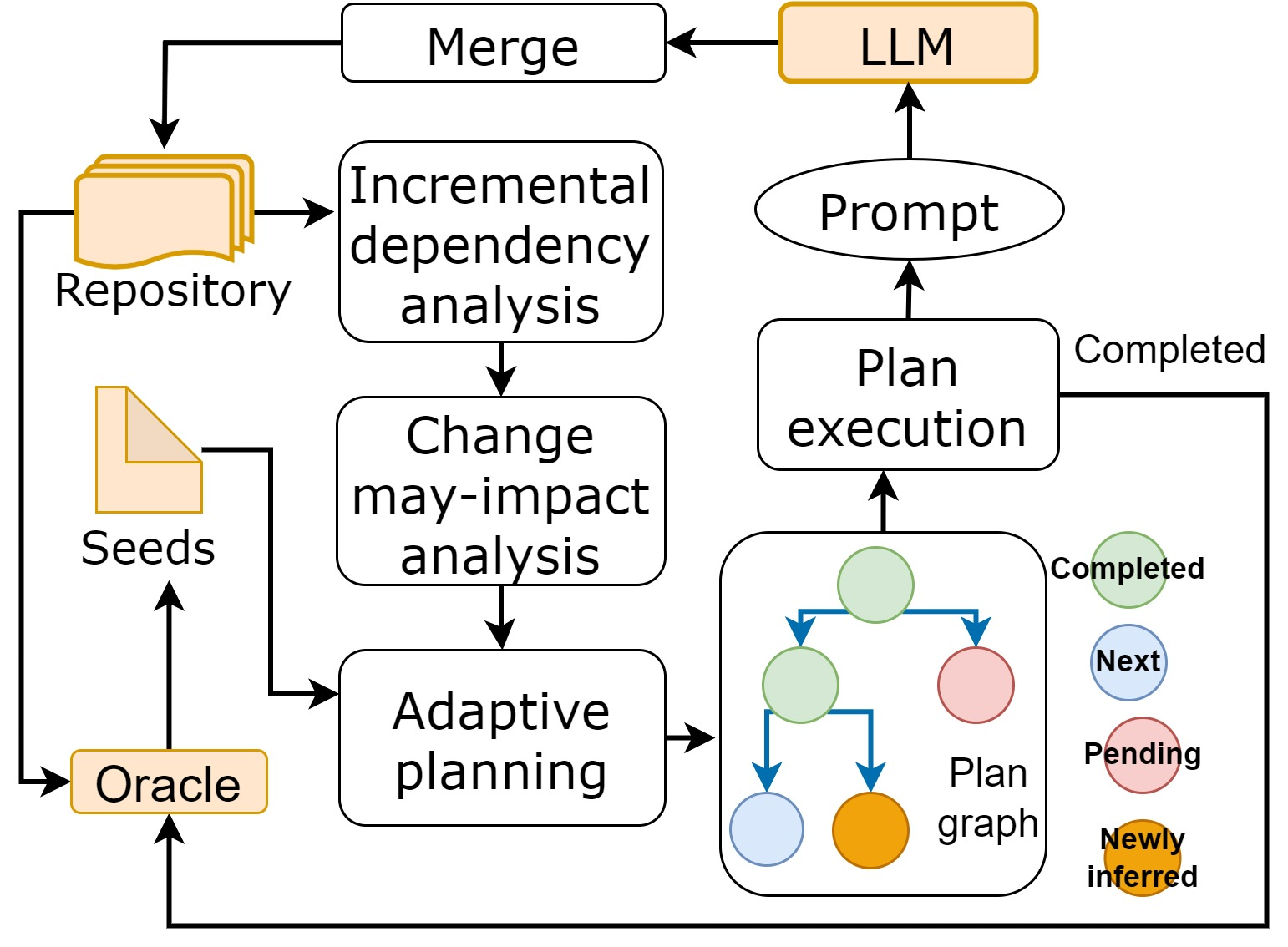}
\vspace*{-7pt}
\caption{Overview of \sysname.}
\label{fig:overview}
\end{minipage}

\end{figure}

The remarkable generative abilities of Large Language Models (LLMs)~\cite{brown2020language,chen_evaluating_2021_Codex,chowdhery2022palm,fried_incoder_2022,openai2023gpt4,touvron2023llama} have opened new ways to automate coding tasks. 
Tools built on LLMs, such as Amazon Code Whisperer~\cite{CodeWhisperer}, GitHub Copilot~\cite{GihubCopilot} and Replit~\cite{Replit}, are now widely used to complete code 

given a natural language intent and context of surrounding code, and also to perform code edits based on natural language instructions~\cite{CopilotChatMode}.
Such edits are typically done for small regions of code such as completing or editing the current line, or the body of the entire method.

While these tools help with the "inner loop" of software engineering where the developer is coding in the editor and editing a small region of code, there are several tasks in the "outer loop" of software engineering that involve the entire code repository. For example, if our code repository uses a library $L$, and the API of library $L$ changes from version $v_n$ to version $v_{n+1}$, we need to migrate our code repository to correctly invoke the revised version. Such a migration task involves making edits not only to all the regions of repository that make calls to the relevant APIs in library $L$, 
but also to regions of the repository (across file boundaries) having transitive syntactic and semantic dependencies on the updated code.

\begin{figure}
\begin{minipage}{\linewidth}
\begin{tabular}{|l@{\!\!\!}|l|}
\hline
\begin{minipage}{0.55\linewidth}
\begin{lstlisting}
tuple<tuple<float, float>, dict> func(float a, float b) {
  string timestamp = GetTimestamp(DateTime.Now);
  |\color{Red} var c = (create\_complex(a,b), new Dictionary<string, string>(){{"time", timestamp}});|
  return c;
}
\end{lstlisting}
\end{minipage}
&
\begin{minipage}{0.45\linewidth}
\begin{lstlisting}
|\color{Green}Complex func(float a, float b)| {
  String timestamp = GetTimestamp(DataTime.Now);
  |\color{Green}dict\_metadata = new Dictionary<string, string>()\{{"time", timestamp\}};|
  |\color{Green}Complex c = create\_complex(a, b, metadata);|
  return c;
}
\end{lstlisting}
\end{minipage}
\\
(a) \code{Create.cs} - Original
&
(b) \code{Create.cs} - Modified (seed edit)
\\
\hline
\begin{minipage}{0.55\linewidth}
\begin{lstlisting}
void process(float a, float b, float k) {
  |\color{Red}var c = func(a, b);|
  |\color{Red}Console.WriteLine(c[0][0], c[0][1]);|
  float norm = compute_norm(c[0][0], c[0][1]);
  Console.WriteLine(norm * k);
}
\end{lstlisting}    
\end{minipage}
&
\begin{minipage}{0.45\linewidth}
\begin{lstlisting}
void process(float a, float b, float k) {
  |\color{Green}Complex c = func(a, b);|
  |\color{Green}Console.WriteLine(c.real, c.imag);|
  float norm = compute_norm(c.real, c.imag);
  Console.WriteLine(norm * k);
}
\end{lstlisting}
\end{minipage}
\\
(c) \code{Process.cs} - Original
&
(d) \code{Process.cs} - Modified (derived edit)
\\
\hline
\end{tabular}
\vspace*{-7pt}
\caption{Relevant code snippets from our repository.}
\label{fig:relevant-coderepository}
\end{minipage}
\end{figure}

This is illustrated in Figure~\ref{fig:complexnumber}, which shows a change in the API for a Complex Numbers library. Our task is to migrate our code repository in accordance with this change. The left side of 
Figure~\ref{fig:relevant-coderepository} shows relevant parts of our code repository that use the Complex Numbers library.
Specifically, the file \code{Create.cs} has the method \code{func}, which invokes the 
\code{create\_complex} method from the library, and \code{Process.cs} has the method
\code{process} which invokes \code{func}.

We can pass the task description from Figure~\ref{fig:complexnumber} and the body of \code{func} to an LLM to generate the revised code for 
\code{func} as shown in the right side of Figure~\ref{fig:relevant-coderepository}. As seen, the LLM has correctly edited the invocation to the \code{create\_complex} API so that it returns an object of type \code{Complex} instead of a tuple of two floating point values.
Note that this edit has resulted in a change to the signature of the method \code{func} -- it now returns an object of type \code{Complex}.
This necessitates changes to callers of method \code{func} such 
 as the \code{process} method in file \code{Process.cs}, shown in the left-bottom of Figure~\ref{fig:relevant-coderepository}. Without a suitable change to the body of the \code{process} method, our code does not build! A suitable change to the \code{process} method which gets the repository to a consistent  state, so that it builds without errors, is shown in the bottom-right of Figure~\ref{fig:relevant-coderepository}.

\myparagraph{Problem Formulation}
The migration task above is representative of a family of tasks that involve  editing an entire code repository for various purposes such as fixing error reports from static analysis or testing, fixing a buggy coding pattern, refactoring, or adding type annotations or other specifications. Each of  these tasks involves a set of {\em seed specifications} such as the one shown in Figure~\ref{fig:complexnumber}, which are starting points for the code editing task. These seed specifications typically trigger other editing requirements on code, and such requirements need to be propagated across dependencies in the code repository to perform other edits across the repository to complete the coding task. Typically, such propagation of edits across dependencies is done manually.

Our goal is to construct a repository-level coding system, which automatically generates {\em derived specifications} for edits such as one required for the \code{process} method in Figure~\ref{fig:relevant-coderepository}, in order to get the repository to a {\em valid} state. Here, validity is defined with respect to an oracle, which can be instantiated to various ways of enforcing repository-level correctness conditions such as building without errors, passing static analysis, passing a type system or a set of tests, or passing a verification tool. 
We define an LLM-driven repository-level coding task as follows:

\begin{tcolorbox}[colback=blue!5,title=LLM-driven Repository-level Coding Task,left=-1pt, right=-7pt, size=small]
Given 
a start state of a repository $R_{start}$, 
a set of seed edit specifications $\Delta_{seeds}$,
an oracle $\Theta$
such that $\Theta(R_{start}) = \ttt$, 
and an LLM $L$, the goal of an \textbf{LLM-driven repository-level coding task} is to reach a repository state $R_{target} = ExecuteEdits(L, R_{start}, P)$ where $P$ is a chain of edit specifications from $\Delta_{seeds} \cup \Delta_{derived}$ where
$\Delta_{derived}$ is a set of derived edit specifications so that
$\Theta(R_{target}) = \ttt$.
\end{tcolorbox}

\mcomment{
\begin{tcolorbox}[colback=blue!5,title=Repository-level editing problem,left=-1pt, right=-7pt, size=small]
Given a start state of a repository $R_{start}$ and a task description $task$, the \textbf{goal of repository level editing} is to reach a repository state $R_{target} = Execute(R_{start}, plan)$ where $plan$ is a sequence of edit specifications from $\Delta_{task} \cup \Delta_{derived}$ such that $\Delta_{task}$ is a set of code change specifications to satisfy the task description and $\Delta_{derived}$ is a minimal set of code change specifications needed to \todo{ensure that the code repository builds.}
\end{tcolorbox}
}

\myparagraph{Proposed Solution}
In this paper, we propose a method to compute derived specifications by framing (LLM-driven) repository-level coding as a {\em planning problem}.
Automated planning~\cite{ghallab2004automated,russell2010artificial} aims to solve multi-step problems, where each step executes one action among many alternatives towards reaching a target state.
It is used in a wide range of areas such as motion planning~\cite{la2011motion}, autonomous driving~\cite{gonzalez2015review}, robotics~\cite{karpas2020automated} and theorem proving~\cite{bundy1988use}.

We present a task-agnostic framework, called \sysname, which synthesizes a multi-step plan to solve the repository-level coding task.
As shown in Figure~\ref{fig:overview},
the input to \sysname\ is a repository, a task with seed specifications expressed through a natural language instruction or a set of initial code edits, a correctness oracle and an LLM.
\sysname\ constructs a \emph{plan graph} where each node in the graph identifies a code edit obligation that the LLM needs to discharge and an edge
indicates that the target node needs to be discharged consequent to the source node. 
\sysname\ monitors the code edits and adaptively extends the plan graph. The edits $\Delta_{seeds}$ follow from the task description, whereas the edits $\Delta_{derived}$ are identified and contextualized based on a novel combination of an incremental dependency analysis, a change may-impact analysis and an adaptive planning algorithm.
The merge block merges the code generated by the LLM into the repository.
Once all the steps in a plan are completed, the repository is analyzed by the oracle.
The task is completed if the oracle validates the repository.
If it finds errors, the error reports are used as seed specifications for the next round of plan generation and execution.

Consider again, the example API migration task specified in Figure~\ref{fig:complexnumber} on code in Figure~\ref{fig:relevant-coderepository}. \sysname performs the edit of the method \code{func} using the instruction in Figure~\ref{fig:complexnumber} as a seed specification. By analyzing the code change between Figure~\ref{fig:relevant-coderepository}(a)--(b), it classifies the change as an \emph{escaping change} as it affects signature of method \code{func}. The change may-impact analysis identifies that the caller(s) of \code{func} may be affected and hence, the adaptive planning algorithm uses caller-callee dependencies to infer a derived specification to edit the method \code{process}, which invokes \code{func}. Both the seed and derived changes are executed by creating suitable prompts for an LLM and the resulting code repository passes the oracle, i.e., builds without errors. Note that this is a simple example with only one-hop change propagation. In practice, the derived changes can themselves necessitate other changes transitively and \sysname handles such cases.

A simpler alternative to our planning is to use the oracle to infer derived specifications. For example, the build system can find the error in the \code{process} method after the seed change is made in Figure~\ref{fig:relevant-coderepository}. This has important limitations. First, not all changes induce build errors even though they result in behavioral changes, e.g., changing the return value from \code{True} to \code{False} without changing the return type. 
Second, the build system is agnostic to cause-effect relationship when code breaks. For example, if the signature of an overriding method is changed as per the seed specification then a similar change is needed in the corresponding virtual method. However, the build system (when run on the intermediate, inconsistent snapshot of the repository) blames the overriding method for not conforming to the virtual method. Na\"ively trying to fix the build error would end up reverting the seed change. 
The static analysis and planning components of \sysname overcome these limitations.
We experimentally compare \sysname against a baseline that uses a build system to iteratively identify breaking changes and uses an LLM to fix them. Our quantitative and qualitative results show that \sysname is superior to this kind of oracle-guided repair technique.

\myparagraph{Contributions}
To the best of our knowledge, the problem of monitoring the effects of code edits made by an LLM to a repository and systematically planning \emph{a chain of inter-dependent edits} has not been identified and solved before.

In the space of repository-level coding tasks, two types of contexts have been found to be useful for prompting LLMs:
(1) \emph{spatial context} to provide cross-file information to the model using static analysis~\cite{pashakhanloo2022codetrek,shrivastava_repository-level_2022_rlpg,ding_cocomic_2022,wei2023typet5,pei_better_2023_awslsp,agrawal2023guiding,shrivastava2023repofusion,liu2023repobench}
or retrieval~\cite{xu2021capturing,zhang_repocoder_2023_repocoder},
and 
(2) \emph{temporal context} to condition the predictions on the history of edits to the repository~\cite{c3po,editpro,gupta2023grace,wei2023coeditor}.
Since \sysname monitors the code changes and maintains a repository-wide dependency graph, we provide both these forms of contexts in a unified framework.
The existing techniques assume that the next edit location is provided by the developer and do not account for the effect of an edit on the dependent code.
In contrast, by inferring the impact of each change, \sysname propagates the changes to dependent code, paving a way to automate repository-level coding tasks through chain of edits.

In summary, we make the following contributions in this paper:
\begin{enumerate}
    \item We are the first to formalize the problem of automating repository-level coding tasks using LLMs, which requires analyzing the effects of code changes and propagating them across the repository. There are currently no systematic and scalable solutions to this problem.
    \item We frame repository-level coding as a planning problem and design a task-agnostic framework, called \sysname, based on a novel combination of an incremental dependency analysis, a change may-impact analysis and an adaptive planning algorithm. \sysname synthesizes a multi-step chain of edits (plan) to be actuated by an LLM.
    \item We experiment with two repository-level coding tasks using the {\tt gpt-4-32k} model: package migration for C\# repositories and temporal code edits for Python repositories. We compare against baselines that use the oracles (a build system for C\# and a static type checker for Python) for identifying derived edit specifications (in contrast to planning used in \sysname). We use the same contextualization method as \sysname in the baselines.
    \item Our results show that \sysname has better match with the ground truth compared to baselines. \sysname is able to get 5/6 repositories to pass the validity checks, whereas the baselines cannot get any of the repositories to pass them. Except for the 2 proprietary repositories, we will release our data and evaluation scripts at \url{https://aka.ms/CodePlan}.
\end{enumerate}

\section{Design}

In this section, we first give an overview of the \sysname algorithm for automating repository-level coding tasks (Section~\ref{sec:algo}).
We then present the static analysis (Section~\ref{sec:static-analysis}) and the adaptive planning and plan execution (Section~\ref{sec:plan-execution})
components of \sysname.

\subsection{The \sysname Algorithm}
\label{sec:algo}

\begin{algorithm}[h]
\begin{lstlisting}[numbers=left]
/* Inputs: R is the source code of a repository, Delta_seeds is a set of seed edit specifications, Theta is an oracle and L is an LLM. */

CodePlan(R, Delta_seeds, Theta, L): 
  let mutable G: |\hlorchid{PlanGraph}| = null in  |\label{line:plan-graph}|
  let mutable D: |\hlcyan{DependencyGraph}| = |\hlcyan{ConstructDependencyGraph}(R)| in |\label{line:dep-graph}|
    while Delta_seeds is not empty |\label{line:loop-begin}|
      IntializePlanGraph(G, Delta_seeds) |\label{line:init-call}|
      AdaptivePlanAndExecute(R, D, G) |\label{line:plan-call}|
      Delta_seeds = Theta(R) |\label{line:loop-end}|

InitializePlanGraph(G, Delta_seeds): |\label{line:init-def}|
  for each |\zug{\text{B, I}}| in Delta_seeds
    AddRoot(G, |\zug{\text{B, I, Pending}}|) |\label{line:init-end}|

AdaptivePlanAndExecute(R, D, G): |\label{line:plan-def}|
  while G has Nodes with Pending status
    let |\zug{\text{B, I, Pending}}| = GetNextPending(G) in
    // |\textbf{First step: extract fragment of code}|
    let Fragmemt = |\hlorchid{ExtractCodeFragment}|(B, R, I) in |\label{line:first-step}|
    // |\textbf{Second step: gather context of the edit}|
    let Context = GatherContext(B, R, D) in  |\label{line:second-step}|
    // |\textbf{Third step: use the LLM to get edited code fragment}|
    let Prompt = |\hlorchid{MakePrompt}|(Fragment, I, Context) in |\label{line:third-step-1}|
    let NewFragment = InvokeLLM(L, Prompt) in |\label{line:third-step-2}|
    // |\textbf{Fourth step: merge the updated code fragment into R}|
    let R = Merge(NewFragment, B, R) in |\label{line:fourth-step-1}|
    let Labels = |\hlcyan{ClassifyChanges}|(Fragment, NewFragment) in |\label{line:fourth-step-2}| 
    let D' = |\hlcyan{UpdateDependencyGraph}\!|(D, Labels, Fragment, NewFragment, B) in |\label{line:fourth-step-3}|
    // |\textbf{Fifth step: adaptively plan and propogate the effect of the edit on dependant code}|
    let BlockRelationPairs |\!\!\!=\!\!| |\hlcyan{GetAffectedBlocks}|(Labels, B, D, D') in |\label{line:fifth-step-1}|
      MarkCompleted(B, G) |\label{line:markcompleted}|
      for each |\zug{\text{B', rel}}| in BlockRelationPairs
        let N = GetNode(B) in
        let M = SelectOrAddNode(B', Nil, Pending) in |\label{line:addnode}|
          AddEdge(G, M, N, rel) |\label{line:fifth-step-2}|
    D := D'

GatherContext(B, R, D): |\label{line:gathercontext-begin}|
  let SC = |\hlorchid{GetSpatialContext}|(B, R) in
  let TC = |\hlorchid{GetTemporalContext}|(G, B) in
    (SC, TC) |\label{line:gathercontext-end}|
\end{lstlisting}
\caption{The \sysname algorithm to automate repository-level coding tasks.
The data structures and functions in \hlcyan{Cyan} and \hlorchid{Orchid} are explained in Section~\ref{sec:static-analysis}--~\ref{sec:plan-execution} respectively.}
\label{algo:codeplan}
\end{algorithm}

The \sysname algorithm (Algorithm~\ref{algo:codeplan}) takes four inputs: (1) the source code of a repository $R$, (2) a set of seed edit specifications for the task in hand, $\Delta_{seeds}$, (3) an oracle, $\Theta$, and (4) an LLM, $L$.

The core data structure maintained by the algorithm is a {\em plan graph} $G$, a directed acyclic graph with multiple root nodes (line~\ref{line:plan-graph}). Each node in the plan graph is a tuple 
$\zug{B, I, Status}$, where $B$ is a block of code (that is, a sequence of code locations) in the repository $R$, $I$ is an edit instruction (along the lines of the example shown in Figure~\ref{fig:complexnumber}), 

and $Status$ is either $pending$ or $completed$.

\begin{figure*}[t]
    \centering
    \scalebox{1}{
    \includegraphics[width=0.95\textwidth]{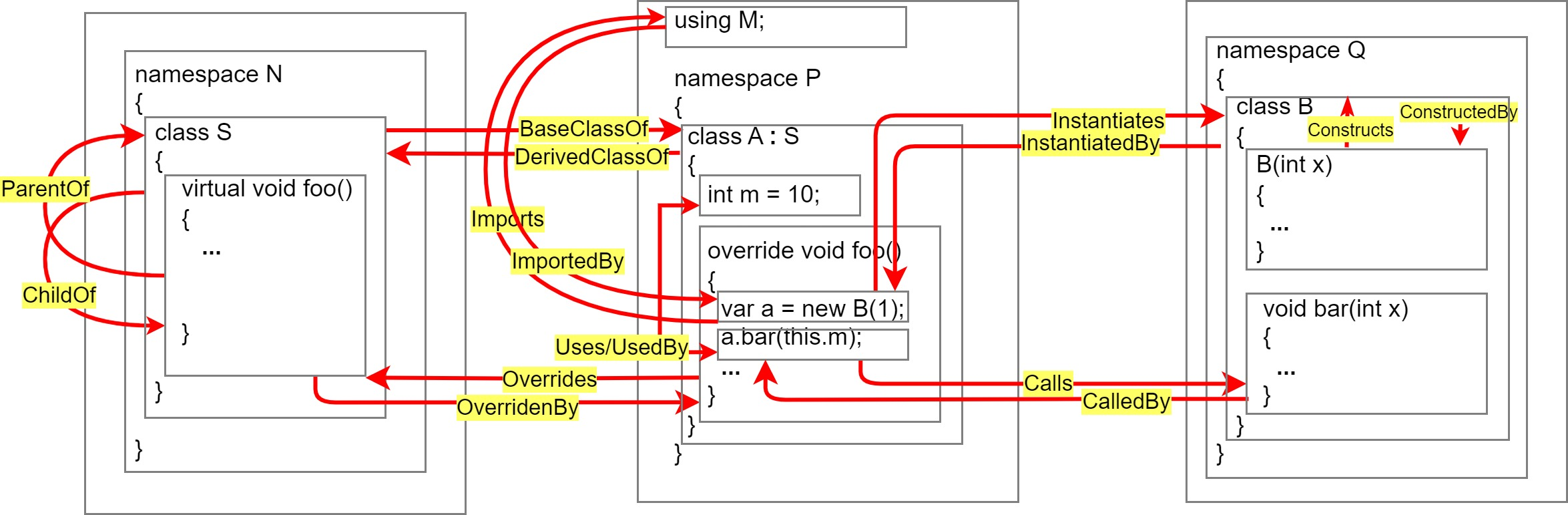}}
    \vspace*{-10pt}
    \caption{Illustration of the dependency graph annotated with relations as the edge labels.}
    \label{fig:dependency-graph}
\end{figure*}

The \sysname algorithm also maintains a \emph{dependency graph} $D$ (line~\ref{line:dep-graph}). Figure~\ref{fig:dependency-graph} illustrates the dependency graph structure. We will discuss it in details in Section~\ref{sec:incremental}. For now, it suffices to know that the dependency graph $D$ represents the syntactic and semantic dependency relations between code blocks in the repository $R$.

The loop at lines~\ref{line:loop-begin}--\ref{line:loop-end} is executed until $\Delta_{seeds}$ is non-empty. Line~\ref{line:init-call} calls the \code{InitializePlanGraph} function (lines~\ref{line:init-def}--\ref{line:init-end}) that adds all the changes in $\Delta_{seeds}$ as root nodes of the plan graph. Each edit specification comprises of a code block $B$ and an edit instruction $I$. 

The status is set to pending for the root nodes (line~\ref{line:init-end}). 
The function \code{AdaptivePlanAndExecute} is called at line~\ref{line:plan-call} which executes the plan, updates the dependency graph with each code change and extends the plan as necessary. Once the plan graph is completely executed, the oracle $\Theta$ is run on the repository. It returns error locations and diagnostic messages which form $\Delta_{seeds}$ for the next round. If the repository passes the oracle's checks then it returns an empty set and the \sysname algorithm terminates.

We now discuss \code{AdaptivePlanAndExecute}, which is the main work horse.
It iteratively picks each pending node and processes it. Processing a pending node with an edit specification for a block $B$ with edit instruction $I$ involves the following five steps:

\begin{enumerate}
\item \textbf{The \emph{first step} (line~\ref{line:first-step}) is to extract the fragment of code to edit.} Simply extracting code of the block $B$ loses information about relationship of $B$ with the surrounding code. Keeping the entire file on the other hand takes up prompt space and is often unnecessary. We found the surrounding context is most helpful when a block belongs to a class. For such blocks, we sketch the enclosing class. That is, in addition to the code of block $B$, we also keep declarations of the enclosing class and its members. As we discuss later, this sketched representation also helps us merge the LLM's output into a source code file more easily.

\item \textbf{The \emph{second step} (line~\ref{line:second-step}) is to gather the context of the edit.} The context of the edit (line~\ref{line:gathercontext-begin}--\ref{line:gathercontext-end}) consists of (a) \emph{spatial context}, which contains related code such as methods called from the block $B$, and (b) \emph{temporal context}, which contains the previous edits that \emph{caused} the need to edit the block $B$. The temporal context is formed by edits along the paths from the root nodes of the plan graph to $B$.

\item \textbf{The \emph{third step} (lines~\ref{line:third-step-1}--\ref{line:third-step-2}) constructs a prompt} for the edit using the fragment extracted in the first step, the instruction $I$ from the edit specification and the context extracted in the second step, and \textbf{invokes the LLM using the prompt} to get the edited code fragment.

\item \textbf{The \emph{fourth step} (lines~\ref{line:fourth-step-1}--\ref{line:fourth-step-3}) merges the edited code back into the repository.} Since the code is updated, many dependency relationships such as caller-callee, class hierarchy, etc. may need to change, and hence, this step also updates the dependency graph $D$. 

\item \textbf{The \emph{fifth and final step} (lines~\ref{line:fifth-step-1}--\ref{line:fifth-step-2}) does adaptive planning to propagate the effects of the current edit on dependant code blocks.} This involves classifying the change in the edited block, and depending on the type of change, picking the right dependencies in the dependency graph to traverse and locate affected blocks. For instance, if the edit of a method $m$ in the current block $B$ involves update to the signature of the method, then all callers of $m$ get affected (the scenario in Figure~\ref{fig:relevant-coderepository}).
For each affected block $B'$ and the dependency relation \mylstinline{rel} connecting $B$ to $B'$ in the dependency graph, we get a pair $\zug{B',\mylstinline{rel}}$. If a node exists for $B'$ in the plan graph and it is pending, then we add an edge from $B$ to $B'$ labeled with \mylstinline{rel} to the plan graph.
Otherwise, the edge is added to a newly created node for $B'$ (line~\ref{line:addnode}).
The block $B$ is marked as completed (line~\ref{line:markcompleted}).
\end{enumerate}

\subsection{Static Analysis Components}
\label{sec:static-analysis}

We now turn our attention to the static analysis components used in \sysname. We will cover all the data structures and functions in \hlcyan{Cyan} background from Algorithm~\ref{algo:codeplan}.

\subsubsection{Incremental Dependency Analysis}
\label{sec:incremental}

An LLM can be provided a code fragment and an instruction to edit it in a prompt. While the LLM may perform the desired edit accurately, analyzing the impact of the edit on the rest of the repository is outside the scope of the LLM call. We believe static analysis is well-suited to do this and propose an incremental dependency analysis for the same.

\cyanparagraph{DependencyGraph}.
Dependency analysis~\cite{aho2007compilers} is used for tracking syntactic and semantic relations between code elements. In our case, we are interested in relations between import statements, methods, classes, field declarations and statements (excluding those that operate only on variables defined locally within the enclosing method). Formally, a \emph{dependency graph} D $= (N,E)$ where $N$ is a set of nodes representing the code blocks mentioned above and $E$ is a set of labeled edges where the edge label gives the relation between the source and target nodes of the edge. Figure~\ref{fig:dependency-graph} illustrates all the relations we track as labeled edges. The relations include 
(1) \emph{syntactic relations} (ParentOf and ChildOf, Construct and ConstructedBy) between a block $c$ and the block $p$ that encloses $c$ syntactically; a special case being a constructor and its enclosing class related by Construct and ConstructedBy,
(2) \emph{import relations} (Imports and ImportedBy) between an import statement and statements that use the imported modules,
(3) \emph{inheritance relations} (BaseClassOf and DerivedClassOf) between a class and its superclass,
(4) \emph{method override relations} (Overrides and OverridenBy) between an overriding method and the overriden method,
(5) \emph{method invocation relations} (Calls and CalledBy) between a statement and the method it calls,
(6) \emph{object instantiation relations} (Instantiates and InstantiatedBy) between a statement and the constructor of the object it creates, and
(7) \emph{field use relations} (Uses and UsedBy) between a statement and the declaration of a field it uses.

\cyanparagraph{ConstructDependencyGraph}.
The dependency relations are derived across the source code spread over the repository through static analysis. We represent the source code of a repository as a forest of abstract syntax trees (ASTs) and add the dependency edges between AST sub-trees. A file-local analysis derives the syntactic and import relations. All other relations require an inter-class, inter-procedural analysis that can span file boundaries. In particular, we use the class hierarchy analysis~\cite{dean1995optimization} for deriving the semantic relations.

\begin{table*}
\centering
\scalebox{0.9}{
\begin{tabular}{p{.15\linewidth}p{.05\linewidth}p{0.4\linewidth}p{.4\linewidth}}
\textbf{Atomic Change} & \textbf{Label} & \textbf{Dependency Graph Update} & \textbf{Change May-Impact Analysis}\\
\hline
\rowcolor{Dandelion}
\multicolumn{4}{c}{Modification Changes}\\
\hline
Body of method M & MMB & 
Recompute the edges incident on the statements in the method body.

&
If an escaping object is modified then Rel(D, M, CalledBy) else Nil.

\\
\rowcolor{Gray}
Signature of method M & MMS &
Recompute the edges incident on the method.

&
Rel(D, M, CalledBy), 
Rel(D, M, Overrides), 
Rel(D, M, OverriddenBy),
Rel(\Dprime, M, Overrides), 
Rel(\Dprime, M, OverriddenBy)
\\
Field F in class C & MF &
Recompute the edges incident on the field.

&
Rel(D, F, UsedBy),
Rel(D, C, ConstructedBy),
Rel(D, C, BaseClassOf), 
Rel(D, C, DerivedClassOf)
\\
\rowcolor{Gray}
Declaration of class C & MC &
Recompute the edges incident on the class.

&
Rel(D, C, InstantiatedBy), 
Rel(D, C, BaseClassOf), Rel(D, C, DerivedClassOf),
Rel(\Dprime, C, BaseClassOf), Rel(\Dprime, C, DerivedClassOf)
\\
Signature of constructor of class C & MCC &
No change.

&
Rel(D, C, InstantiatedBy), Rel(D, C, BaseClassOf), Rel(D, C, DerivedClassOf)
\\
\rowcolor{Gray}
Import/Using statement I & MI &
Recompute the edges incident on the import statement.

&
Rel(D, I, ImportedBy)
\\

\hline

\rowcolor{Dandelion}
\multicolumn{4}{c}{Addition Changes}\\
\hline

Method M in class C & AM &
Add new node and edges by analyzing the method.
If C.M overrides a base class method B.M then redirect the Calls/CalledBy
edges from B.M to C.M if the receiver object is of type C.

&
Rel(D, C, BaseClassOf), Rel(D, C, DerivedClassOf), 
Rel(\Dprime, M, CalledBy)
\\
\rowcolor{Gray}
Field F in class C & AF &
Add new node and edges by analyzing the field declaration.

&
Rel(D, C, ConstructedBy), Rel(D, C, BaseClassOf), Rel(D, C, DerivedClassOf)
\\
Declaration of class C & AC & 
Add new node and edges by analyzing the class declaration.

& 
Nil
\\
\rowcolor{Gray}
Constructor of class C & ACC &
Add new node and edges by analyzing the constructor.

&
Rel(D, C, InstantiatedBy), Rel(D, C, BaseClassOf), Rel(D, C, DerivedClassOf)
\\
Import/Using statement I & AI &
Add new node and edges by analyzing the import statement.

&
Nil
\\
\hline

\rowcolor{Dandelion}
\multicolumn{4}{c}{Deletion Changes}\\
\hline
Method M in class C & DM &
Remove the node for M and edges incident on M.
If C.M overrides a base class method B.M then redirect the Calls/CalledBy
edges from C.M to B.M if the receiver object is of type C.

&
Rel(D, M, CalledBy), 
Rel(D, M, Overrides), 
Rel(D, M, OverriddenBy)
\\
\rowcolor{Gray}
Field F in class C & DF &
Remove the node of the field and edges incident on it.

&
Rel(D, F, UsedBy), Rel(D, C, ConstructedBy), Rel(D, C, BaseClassOf), Rel(D, C, DerivedClassOf)
\\
Declaration of class C & DC &
Remove the node of the class and edges incident on it.

&
Rel(D, C, InstantiatedBy), 
Rel(D, C, BaseClassOf), 
Rel(D, C, DerivedClassOf)
\\
\rowcolor{Gray}
Constructor of class C & DCC &
Remove the edges incident on the class due to object instatiations using the constructor.

&
Rel(D, C, InstantiatedBy), Rel(D, C, BaseClassOf), Rel(D, C, DerivedClassOf)
\\
Import/Using statement I & DI &
Remove the node of the import statement and edges incident on it.

&
Rel(D, I, ImportedBy)
\\
\bottomrule
\end{tabular}
}
\caption{Rules for updating the dependency graph and for change may-impact analysis for atomic changes. We refer to the dependency graphs before and after the updates by D and \Dprime respectively.}
\label{tab:rules}
\end{table*}

\cyanparagraph{ClassifyChanges}.
As discussed in Section~\ref{sec:algo}, in the fourth step, \sysname merges the code generated by the LLM into the repository. By pattern-matching the code before and after, we classify the code changes. Table~\ref{tab:rules} (the first and second columns) gives the types of atomic changes and their labels. Broadly, the changes are organized as modification, addition and deletion changes, and further by which construct is changed. We distinguish between method body and method signature changes. Similarly, we distinguish between changes to a class declaration, to its constructor or to its fields. The changes to import statements or the statements that use imports are also identified. These are \emph{atomic changes}. An LLM can make multiple simultaneous edits in the given code fragment, resulting in multiple atomic changes, all of which are identified by the \lstinline{ClassifyChanges} function.

\cyanparagraph{UpdateDependencyGraph}.
As code generated by the LLM is merged, the dependency relations associated with the code at the change site are re-analyzed. Table~\ref{tab:rules} (the third column) gives the rules to update the dependency graph D to \Dprime based on the labels inferred by \lstinline{ClassifyChanges}. For modification changes, we recompute the relations of the changed code except for constructors. A constructor is related to its enclosing class by a syntactic relation which does not have to be recomputed. For addition changes, new nodes and edges are created for the added code. Edges corresponding to syntactic relations are created in a straightforward manner. If a change simultaneously adds an element (an import, a method, a field or a class) and its uses, we create a node for the added element before analyzing the statements that use it. Addition of a method needs special handling as shown in the table: if an overriding method C.M is added then the Calls/CalledBy edges incident on the matching overriden method B.M are redirected to C.M if the call is issued on a receiver object of type C. The deletion of an overriding method requires an analogous treatment as stated in Table~\ref{tab:rules}. All other deletion changes require removing nodes and edges as stated in the table.

\subsubsection{Change May-Impact Analysis}
\label{sec:impact}
In the fifth step, \sysname identifies the code blocks that may have been impacted by the code change by the LLM.
Let Rel(D, B, rel) be the set of blocks that are connected to a block B via relation rel in the dependency graph D.
Let D and \Dprime be the dependency graph before and after the updates in Table~\ref{tab:rules}.

\cyanparagraph{GetAffectedBlocks}.
The last column in Table~\ref{tab:rules} tells us how to identify blocks affected by a code change for each type of change.
When the body of a method M is edited, we perform escape analysis~\cite{choi1999escape,blanchet2003escape} to identify if any object accessible in the callers of M (an escaping object) has been affected by the change. If yes, the callers of M (identified through Rel(D, M, CalledBy)) are identified as affected blocks. Otherwise, the change is localized to the method and there are no affected blocks. If the signature of a method is edited, the callers and methods related to it through method-override relation in the inheritance hierarchy are affected. The signature change itself can affect the Overrides and OverridenBy relations, e.g., addition or deletion of the \code{@Override} access modifier. Therefore, the blocks related by these relations in the updated dependency graph \Dprime are also considered as affected as shown in Table~\ref{tab:rules} (the row with MMS label). When a field F of a class C is modified, the statements that use F, the constructors of C and sub/super-classes of C are affected.
When a class is modified, the methods that instantiate it and its sub/super-classes as per D and \Dprime are affected.
A modification to a constructor has a similar rule except that such a change does not change inheritance relations and hence, only D is required.
When an import statement I is modified, the statements that use the imported module are affected.

The addition and deletion changes are less complex than the modification changes, and their rules are designed along the same lines as discussed above. In the interest of space, we do not explain each of them step-by-step. We assume that there is no use of a newly added class or an import in the code. Therefore, adding them does not result in any affected blocks. In our experiments, we have found the rules in Table~\ref{tab:rules} to be adequate. However, \sysname can be easily configured to accommodate variations of the rules in Table~\ref{tab:rules} if necessary.

\subsection{Adaptive Planning and Plan Execution}
\label{sec:plan-execution}

We now discuss the data structures and functions from Algorithm~\ref{algo:codeplan} in the \hlorchid{Orchid} background.

\subsubsection{Adaptive Planning}
Having identified the affected blocks (using \lstinline{GetAffectedBlocks}), \sysname creates change obligations that need to be discharged using an LLM to make the dependent code consistent with the change. As discussed in Section~\ref{sec:algo}, this is an iterative process.

\orchidparagraph{PlanGraph}.
A \emph{plan graph} P $= (O, C)$ is a directed acyclic graph with a set of \emph{obligations} $O$, each of which is a triple \zug{B, I, status} where B is a block, I is an instruction and status is either pending or completed. An edge in $C$ records the \emph{cause}, the dependency relation between the blocks in the source and target obligations. In other words, the edge label identifies which Rel clause in a change may-impact rule in Table~\ref{tab:rules} results in creation of the target obligation.

\orchidparagraph{ExtractCodeFragment}.
As discussed in the first step in Section~\ref{sec:algo}, simply extracting code for a block B is sub-optimal as it loses context. The \lstinline{ExtractCodeFragment} function takes the whole class the code block belongs to, keeps the complete code for B and retains only declarations of the class and other class members. We found this to be useful because the names and types of the class and other members provide additional context to the LLM. Often times the LLM needs to make multiple simultaneous changes. For example, in some of our case studies, the LLM has to add a field declaration, take an argument to a constructor and use it within the constructor to initialize the field. Providing the sketch of the surrounding code as a code fragment to the LLM allows the LLM to make these changes at the right places. The code fragment extraction logic is implemented by traversing the AST and "folding" away the subtrees (e.g., method bodies) that are sketched. As stated in Section~\ref{algo:codeplan}, this sketched representation also allows us to place the LLM generated code back into the AST without ambiguity, even when there are multiple simultaneous changes.

\orchidparagraph{GetSpatialContext}.
Spatial context in CodePlan refers to the arrangement and relationships of code blocks within a codebase, helping understand how classes, functions, variables, and modules are structured and interact. It's crucial for making accurate code changes. CodePlan utilizes the dependency graph to extract spatial context, representing code as nodes and their relationships as edges. This graph enables CodePlan to navigate codebases, identify relevant code blocks, and maintain awareness of their spatial context. As a result, when generating code edits, the dependency graph empowers CodePlan to make context-aware code modifications that are consistent with the code's spatial organization, enhancing the accuracy and reliability of its code editing capabilities.

\orchidparagraph{GetTemporalContext}.
The plan graph records all change obligations and their inter-dependences. Extracting temporal context is accomplished by linearizing all paths from the root nodes of the plan graph to the target node. Each change is a pair of the code fragments before and after the change. The temporal context also states the "causes" (recorded as edge labels) that connect the target node with its predecessor nodes. For example, if a node A is connected to B with a CalledBy edge, then the temporal context for B is the before/after fragments for A and a statement that says that "B calls A", which helps the LLM understand the cause-effect relation between the latest temporal change (change to A) and the current obligation (to make a change to B).

\subsubsection{Plan Execution}
\sysname iteratively selects a pending node in the plan graph and invokes an LLM to discharge the change obligation.

\orchidparagraph{MakePrompt}.
Having extracted the code fragment to be edited along with the relevant spatial and temporal context, we construct a prompt to pass to the LLM with the structure given below. We open with the task specific instructions \circled{$\textrm{p}_1$} followed by listing the edits made in the repository so far \circled{$\textrm{p}_2$} that are relevant to the fragment being edited. The next section \circled{$\textrm{p}_3$} notes how each of the fragments present in \circled{$\textrm{p}_2$} are related to the fragment to be edited. This is followed by the spatial context \circled{$\textrm{p}_4$} and the fragment to the edited \circled{$\textrm{p}_5$}.

\prompt{Prompt Template}{
\circled{$\textrm{p}_1$} Task Instructions: {\it Your task is to} $\hdots$
\newline

\circled{$\textrm{p}_2$} Earlier Code Changes (Temporal Context): {\it These are edits that have been made in the code-base previously -}
\newline

\ \ {\tt Edit 1: \\
\ \ \ \ Before: <<code\_before>> \\
\ \ \ \ After: <<code\_after>> \\
}
$\ \ \ \cdots$
\newline

\circled{$\textrm{p}_3$} Causes for Change: {\it The change is required due to -}
\newline

\ \ {\tt <<code\_to\_be\_edited>> } is related to {\tt <<code\_changed\_earlier>>} by {\tt <<cause>>} \\
$\ \ \ \cdots$
\newline

\circled{$\textrm{p}_4$} Related Code (Spatial Context): {\it The following code maybe related -}
\newline

\ \ {\tt <<related\_code\_block-1>> } \\
$\ \ \ \cdots$
\newline

\circled{$\textrm{p}_5$} Code to be Changed Next: {\it The existing code is given below -}
\newline

\ \ {\tt <<code\_to\_be\_edited>> }
\newline

{\it Edit the "Code to be Changed Next" and produce "Changed Code" below. Edit the "Code to be Changed Next" according to the "Task Instructions" to make it consistent with the "Earlier Code Changes", "Causes for Change" and "Related Code".
If no changes are needed, output "No changes." }

}

\myparagraph{Oracle and Plan Iterations}
Once all the nodes in the plan graph are marked as completed and no new nodes are added, an \emph{iteration} of repository-level code edits is completed. As shown in Figure~\ref{fig:overview}, the oracle is invoked on the repository. If the oracle flags any errors (e.g., build errors), the error locations and diagnostic messages are added as seed changes for the next iteration and the adaptive planning resumes once again. If the oracle does not flag any errors, \sysname terminates.

\section{Implementation}

In this section, we provide a detailed overview of the implementation components that constitute the core of our method.

\begin{figure*}[t]
    \centering
    \scalebox{1}{
    \includegraphics[width=1.0\textwidth]{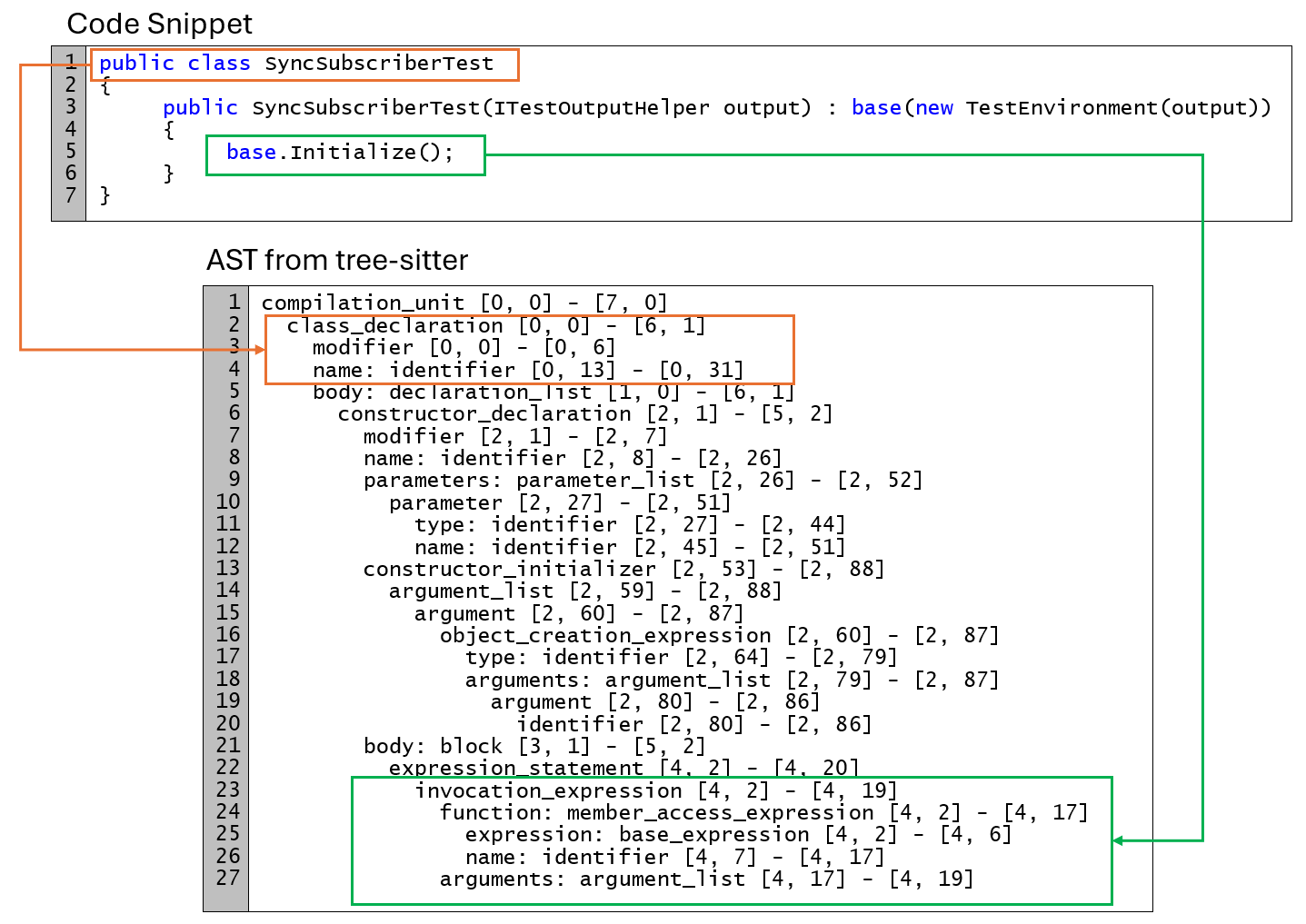}}
    \vspace*{-10pt}
    \caption{AST structure for a C\# code snippet produced by \tt{tree-sitter}.}
    \label{fig:tree_sitter_ast}
\end{figure*}

\myparagraph{Dependency Graph Construction}
At the core of the CodePlan methodology lies the Dependency Graph, which is instrumental in representing the intricate relationships between code blocks. To build this Dependency Graph from a code repository, we adopt a systematic approach. Initially, we parse all the code files within the repository, utilizing the {\tt tree-sitter} library~\cite{tree_sitter} to generate an AST-like structure. This structured representation simplifies the identification of various fundamental code blocks within the codebase. For instance, Figure \ref{fig:tree_sitter_ast} exemplifies an AST structure for a C\# code snippet produced by {\tt tree-sitter}. Code blocks are identified at different levels, including Classes, Methods, import statements, and non-class expressions. For instance, in Figure~\ref{fig:tree_sitter_ast}, the subtree rooted at the \code{class\_declaration} node corresponds to the \code{SyncSubscriberTest} class.

\myparagraph{Relation Identification in C\#}
In the context of C\# repositories, the establishment of edges within the Dependency Graph involves the careful tracing of relationships within the AST. We have devised custom logic for each type of relationship outlined in Figure \ref{fig:dependency-graph}, encompassing vital connections such as caller-callee, overrides-overridden, base class-derived class, and others. To illustrate, for the Caller/Callee relationship, we search for \code{invocation\_expression} nodes within the AST. Subsequently, we process the sub-tree beneath these nodes to resolve essential details such as the target class and the invoked method's name. Armed with this information, we create Calls/CalledBy relation links between the code block initiating the method call and the corresponding method block within the target class. While we have implemented custom logic for these relations, it's important to note that alternative dependency analysis tools for C\# such as Language Servers for C\# (LSP)~\cite{OmniSharp}, CodeQL~\cite{CodeQL}, or similar solutions can also be integrated into our system, owing to its inherent flexibility.

\myparagraph{Relation Identification in Python}
For Python repositories, we use Jedi~\cite{Jedi} - a static analysis tool which discovers references and declarations of symbols throughout the codebase. These capabilities are harnessed to identify edges in the Dependency Graph for relationships such caller-callee, overrides-overridden, and base class-derived class.

\myparagraph{Integration of GPT-4 for Code Edits}
CodePlan leverages the remarkable capabilities of GPT-4~\cite{openai2023gpt4} to perform code edits effectively. During the construction of input data for the edit model, we meticulously provide temporal context, spatial context, and the actual code to be edited in the form of code snippets. These code snippets represent classes or methods that contain the edit site and are meticulously structured in a sketched representation, as stated in Section~\ref{sec:algo}. This sketched representation ensures that the model is enriched with a substantial context for each edit site, significantly enhancing the quality and accuracy of the edits it generates.
 
\myparagraph{Language Extensibility}
While our current implementation proficiently supports C\# and Python repositories, extending support to repositories in other programming languages is a straightforward endeavor. It primarily entails creating a dependency graph with the relations identified in Figure \ref{fig:dependency-graph} and incorporating it into the CodePlan framework, thereby allowing for seamless adaptation to a diverse array of programming languages.

\section{Experimental Design}
In this section, we'll explain how we conducted our experiments to test \sysname. We'll start by talking about the different sets of data we used. Then, we'll discuss the methods we compared \sysname to, which are like our reference points. Finally, we'll explain how we measured the results to see how well \sysname performed compared to the other methods. 

\subsection{Datasets} \label{sec:datasets}

In our experiments, we utilized diverse datasets representing a wide spectrum of code repositories with varying complexities and sizes. These datasets allowed us to thoroughly evaluate the performance of \sysname in different real-world scenarios.

\myparagraph{Internal Repositories (Int-1 and Int-2)} These repositories are proprietary and belong to a large product company. They are characterized by their large size, complex patterns, and are typical of production-level codebases. The primary task we focused on here was the migration of these repositories from a legacy logging framework to a modern logging framework. This migration involved non-trivial changes, including creating service-specific loggers using a logging factory, passing the logger through call chains, managing class hierarchies, storing logger references at different scopes (class, method, etc.), and handling loggers at both static and non-static classes/methods. These two production repositories, Int-1 and Int-2, were chosen for their distinct coding styles and design patterns, providing a comprehensive internal dataset for our evaluation.

\myparagraph{External Repositories (Public GitHub)} We also considered external repositories from GitHub to diversify our dataset. These repositories were chosen to represent two different coding tasks: Migration and Temporal Edits (discussed next).

\underline{Migration Task.} This task involves migrating APIs or addressing breaking changes within a codebase. Examples include updating dependencies, adapting to changes in external libraries, or aligning code with new coding standards. It is characterized by its complexity, often requiring consistent updates across numerous code files and dependencies. To select these repositories, we looked for those containing commits and pull requests related to various kinds of migrations (API, frameworks, etc.). We filtered for repositories with at least 50 files.

\underline{Temporal Edits Task.} This task involves orchestrating a sequence of code changes given some initial code-change. Many code-changes can be characterised as temporal edits including refactoring or addition/removal of functionality. A temporal edit task is specified by a set of initial edits (usually made by the user) along with derived edits precipitated by the seed edits. The task for the tool is to infer the derived edits from the set of initial edits. An example temporal edit task can be where the initial edit is adding an argument to a method and the derived edits are changes to all the places where this method is called. We identify temporal edit tasks from commits made in public repositories on GitHub. We consider python repositories with permissive licenses, sort by stars, filter out documentation/tutorial related repositories and from amongst candidates with at least 10,000 stars, we select commits made after 1st November 2021 (after the cut-off date for training data of GPT-4) containing multiple related changes. 

\myparagraph{Source/Target/Predicted Repositories} To collect the code changes for the migration and temporal edit tasks, we obtained files from GitHub before and after the commits. We refer to these as the Source repository (before commit) and Target repository (after commit). Analyzing these changes allowed us to identify seed changes through manual inspection. For instance, in the NUnit to XUnit migration, one of the seed edits involved replacing \code{Console.WriteLine} with writing to an \code{ITestOutputHelper} object. With the Source repository and seed change instructions, \sysname was tasked with making the necessary changes to the Source repository, resulting in what we call the Predicted repository. If \sysname successfully executed all changes, the Predicted repository should match the Target repository, providing a robust evaluation of its capabilities.

\myparagraph{Preprocessing Source and Target Repositories} When dealing with large repositories, it's common for multiple developers to contribute code, resulting in various coding styles driven by individual preferences. For instance, one developer might consistently use the \code{this} qualifier to reference class members, while another may not. When \sysname executes changes through LLM prompting, it tends to establish a uniform style throughout the codebase, which may involve enforcing the consistent use of the \code{this} qualifier, among other things. While these changes ensure functional equivalence, they can impact evaluation metrics when comparing the Predicted repository with the Target repository. To address these issues, we employ a manual preprocessing step on both the Source and Target repositories. This preprocessing aims to establish uniformity across various files within the repository. By doing so, we provide a fair basis for comparing \sysname's performance against the ground truth in these diverse codebases.

\begin{table}[t]
\centering
\begin{tabularx}{\textwidth}{|>{\columncolor{Dandelion}}X |ccc|ccc@{}|}
\hline
\multicolumn{1}{|c|}{\cellcolor{Dandelion}} & \multicolumn{3}{c|}{\cellcolor{Dandelion}\textbf{\begin{tabular}[c]{@{}c@{}}Migration \\ (C\#)\end{tabular}}} & \multicolumn{3}{c|}{\cellcolor{Dandelion}\textbf{\begin{tabular}[c]{@{}c@{}}Temporal Edits\\ (Python)\end{tabular}}} \\ \cline{2-7} 
\multicolumn{1}{|c|}{\multirow{-2}{*}{\cellcolor{Dandelion}\textbf{Repositories}}} & \multicolumn{1}{l|}{\cellcolor{Dandelion}\textbf{Int-1}} & \multicolumn{1}{l|}{\cellcolor{Dandelion}\textbf{Int-2}} & \multicolumn{1}{l|}{\cellcolor{Dandelion}\textbf{Ext-1}} & \multicolumn{1}{l|}{\cellcolor{Dandelion}\textbf{T-1}} & \multicolumn{1}{l|}{\cellcolor{Dandelion}\textbf{T-2}} & \multicolumn{1}{l|}{\cellcolor{Dandelion}\textbf{T-3}} \\ \hline
\textbf{Number of files} & \multicolumn{1}{r|}{91} & \multicolumn{1}{r|}{168} & \multicolumn{1}{r|}{55} & \multicolumn{1}{r|}{21} & \multicolumn{1}{r|}{137} & 4 \\ \hline
\textbf{Lines of code} & \multicolumn{1}{r|}{8853} & \multicolumn{1}{r|}{16476} & \multicolumn{1}{r|}{8868} & \multicolumn{1}{r|}{3883} & \multicolumn{1}{r|}{20413} & 1874 \\ \hline
\textbf{Number of files changed} & \multicolumn{1}{r|}{47} & \multicolumn{1}{r|}{97} & \multicolumn{1}{r|}{21} & \multicolumn{1}{r|}{2} & \multicolumn{1}{r|}{2} & 3 \\ \hline
\textbf{Number of seed changes} & \multicolumn{1}{r|}{41} & \multicolumn{1}{r|}{63} & \multicolumn{1}{r|}{42} & \multicolumn{1}{r|}{2} & \multicolumn{1}{r|}{1} & 1 \\ \hline
\textbf{Number of derived changes} & \multicolumn{1}{r|}{110} & \multicolumn{1}{r|}{375} & \multicolumn{1}{r|}{16} 0 & \multicolumn{1}{r|}{8} & \multicolumn{1}{r|}{3} & 10 \\ \hline
\textbf{Size of diff (lines)} & \multicolumn{1}{r|}{1744} & \multicolumn{1}{r|}{4902} & \multicolumn{1}{r|}{1024} & \multicolumn{1}{r|}{104} & \multicolumn{1}{r|}{15} & 39 \\ \hline
\textbf{Size of seed edits (lines)} & \multicolumn{1}{r|}{242} & \multicolumn{1}{r|}{242} & \multicolumn{1}{r|}{379} & \multicolumn{1}{r|}{76} & \multicolumn{1}{r|}{4} & 1 \\ \hline
\textbf{Prompt template size (lines)} & \multicolumn{1}{r|}{81} & \multicolumn{1}{r|}{81} & \multicolumn{1}{r|}{81} & \multicolumn{1}{r|}{75} & \multicolumn{1}{r|}{75} & 75 \\ \hline
\textbf{URL} & \multicolumn{1}{r|}{-} & \multicolumn{1}{r|}{-} & \multicolumn{1}{r|}{\cite{repo_ext_1}} & \multicolumn{1}{r|}{\cite{whisper}} & \multicolumn{1}{r|}{\cite{audiocraft}} & \cite{JARVIS} \\ \hline
\end{tabularx}
\caption{Dataset statistics. Int-1,2 are internal (proprietary) repositories, Ext-1 and T-1,2,3 are external (public GitHub) repositories.}
\label{tab:dataset_stats}
\end{table}

Table \ref{tab:dataset_stats} summarizes the dataset statistics for both Migration (C\#) and Temporal Edits (Python) repositories, including the names of repositories (both internal and external) and various key statistics including their size, code changes, and other relevant metrics:

\begin{itemize}
    \item \emph{Number of Files}: The total count of files in each repository.
    \item \emph{Lines of Code}: The cumulative number of lines of code across all files.
    \item \emph{Number of Files Changed}: The count of files that have undergone changes between the source and target repositories.
    \item \emph{Number of Seed Changes}: The number of initial edits, often considered as the starting point of code changes.
    \item \emph{Number of Derived Changes}: The count of subsequent edits that follow the initial seed changes.
    \item \emph{Size of Diff}: The number of lines that differ between the source and target repositories.
    \item \emph{Size of Seed Edits}: When the seed edits are made directly on the code, it represents the number of lines of initial edits. When the seed edits are made through LLM instruction, it denotes the size of the instruction text.
    \item \emph{Prompt Template Size}: This number represents the size of the LLM prompt template used by \sysname. The same template is employed for all the migration repository tasks, and another similar template is utilized for all the Temporal Edit repository tasks.
\end{itemize}

These metrics not only offer a comprehensive overview of the dataset characteristics but also highlight the considerable advantages of using \sysname over manual processes, especially for large repositories. In manual scenarios, human effort is required to painstakingly identify dependent changes and implement each modification. Notably, metrics such as "Size of Diff" and "Size of Seed Edits" provide insights into the developer effort involved. \sysname, on the other hand, automates these changes, effectively reducing the burden on developers. Furthermore, it's worth noting that the effort required to craft LLM instructions for \sysname is significantly less than the extensive manual labor required for making all the code changes. These metrics collectively demonstrate the efficiency and effectiveness of \sysname across diverse codebases, emphasizing its potential to streamline development workflows and save valuable developer time.

\subsection{Oracles and Baselines} \label{sec:oracles_baselines}

\myparagraph{Oracles}
Recall that our definition of repository-level coding tasks is centered around satisfying an oracle that can determine the validity of our solution. 
In our experiments, we consider two specific instatiations of such an oracle.
For the C\# migration tasks, we define the oracle as passing the C\# build tools without any errors.
For temporal edits scenario, we use Pyright \cite{pyright}, a static checker for python as the oracle.

\myparagraph{Oracle-Guided Repair} 
Both of these oracles take a codebase as input and can output a list of errors in that codebase. This leads naturally to the formulation of baseline approaches to our tasks which we term as Oracle-Guided Repair.
These are simple reactive approaches where at each step we attempt to rectify the errors flagged by the oracle.
For the C\# migration scenario, the baseline is Build-Repair and for temporal edits it is Pyright-Repair according to the oracles used in both.

The oracle-guide repair works in the following steps:
\begin{enumerate}
\item \emph{Initial Edit}: The process begins with application of the initial seed edit to the codebase.
\item \emph{Build and Error Detection}: After the seed edits, we invoke the oracle which detects errors in the codebase arising from the seed edits.
\item \emph{Error Message Analysis}: The error messages generated by the oracle is then parsed to precisely identify the location of the error within the code.
\item \emph{LLM Patching}: Subsequently, the error message, along with the code fragment from the flagged location, is passed to an LLM. The LLM leverages its code generation capabilities to generate a patch or fix for the identified error. For fair comparison with \sysname, we use our implementation for spatial and temporal context extraction in oracle-guided repair. That is, the key difference between \sysname and oracle-guided repair is that \sysname uses adaptive planning whereas oracle-guided repair uses the diagnostic information generated by the oracle, but they use the same contextualization machinery.

\end{enumerate}

Note that Oracle-Guided Repair approaches are reactive and lack a comprehensive "change may-impact analysis." 
This means that they may not thoroughly assess how the proposed code changes could affect other parts of the codebase. 
As a result, the fixes generated by such approaches may be incomplete or incorrect, especially when dealing with complex coding tasks.

\myparagraph{Alternate Edit Model: Coeditor~\cite{wei2023coeditor}}
\sysname by default leverages the text and code processing abilities to LLMs to make local edits to code fragments given proper context.
However, in theory, the incremental dependancy analysis, change may-impact analysis and adaptive planning components of \sysname can be used in conjunction with any tool or model that can make localized edits to code following some provided intent.
Coeditor \cite{wei2023coeditor} is a transformer based model fine-tuned to edit code snippets while taking into account prior edits made in the repository.
Such a model is good fit for the Temporal Edits task, where we need to make a sequence of edits from a set of seed edits, where each edit is dependant on some subset of the previous edits. Indeed, Coeditor is evaluated on the Temporal Edits task in~\cite{wei2023coeditor}.
In order to demonstrate the generality of our analysis and planning, we evaluate how our approach performs in the temporal edits scenario, when replacing \code{gpt-4-32k} with Coeditor as the edit model.

\subsection{Evaluation Metrics} \label{sec:metrics}

The evaluation metrics employed in our study are aimed at assessing how effectively \sysname (or a baseline) propagates changes across the entire code repository and the correctness of each of these changes. To achieve this, we rely on two key metrics: Block Metrics and Edit Metrics.

\myparagraph{Block Metrics}
Block Metrics help us understand \sysname's ability to accurately identify code blocks in need of modification. These metrics include:
\begin{itemize}
\item \emph{Matched Blocks}: These are code blocks that exist in the Source Repository, have been edited in the Target Repository, and have also been edited in the Predicted Repository. Essentially, these are blocks that \sysname successfully identifies for change.
\item \emph{Missed Blocks}: Missed Blocks refer to code blocks present in the Source Repository that have been edited in the Target Repository but were not edited in the Predicted Repository. In other words, these are blocks that \sysname failed to modify when it should have.
\item \emph{Spurious Blocks}: Spurious Blocks are code blocks found in the Source Repository that were not edited in the Target Repository but were incorrectly edited by \sysname in the Predicted Repository. These represent edits that \sysname made unnecessarily.
\end{itemize}
The ideal outcome is to have a high number of Matched Blocks and low numbers of Missed and Spurious Blocks.

\myparagraph{Edit Metrics} 
While Block Metrics assess the identification of code blocks, Edit Metrics delve into the correctness of the modifications made by \sysname. These metrics include:
\begin{itemize}
\item \emph{Levenshtein Distance}: Levenshtein Distance measures the edit distance at the file level between the Predicted Repository and the Target Repository. It calculates how many changes were made to transform one file into the other. A higher Levenshtein Distance indicates that \sysname did not make the correct changes to the repository.
\item \emph{Diff BLEU}: Typically, we use BLEU~\cite{papineni2002bleu}, a common metric in Natural Language Processing, to measure text similarity. However, when applied in our tasks, BLEU can produce overly high similarity scores because code edits for specific tasks often involve only a small portion of the file. To address this issue, we compute Diff BLEU: a modified version of the BLUE score, denoted as {\tt BLUE(DIFF(Source repo file, Target repo file), DIFF(Source repo file, Predicted repo file))}. Here, {\tt DIFF} calculates the differences (diff hunks) between two files. What sets Diff BLUE apart is its focus on comparing the modified sections of code between the Predicted and Target Repositories while disregarding common code. When the modifications in Predicted and Target Repositories match precisely, Diff BLUE yields a score of 1.0, indicating a high level of correctness and alignment in handling code modifications.
\end{itemize}

In summary, these evaluation metrics provide a comprehensive assessment of \sysname's performance, both in terms of identifying code blocks for modification and ensuring that the modifications made are correct.

\section{Results and Analysis}

In this section, we present empirical results to answer the following research questions:

\begin{enumerate}
    \item[RQ1:] How well \sysname is able to localize and make the required changes to automate repository-level coding tasks?
    \item[RQ2:] How important are the temporal and spatial contexts for \sysname's performance?
    \item[RQ3:] What are the key differentiators that allow \sysname to outperform baselines in solving complex coding tasks?
\end{enumerate}

% \begin{table*}[t]
% \centering
% \begin{lstlisting}[
% basicstyle={\scriptsize\ttfamily},
% identifierstyle={\color{black}},
% numbers=none,
% language=no_lang,
% breaklines=true,
% breakindent=0em,
% frame=single,
% % caption={My Caption},captionpos=b
% ]

% Code snippet
% ```python
% {{ input_code }}
% ```
% \end{lstlisting}
% \captionof{figure}{Prompt template.}
% \label{prompt:zero-shot}
% \end{table*}

\subsection{RQ1: How well \sysname is able to localize and make the required changes to automate repository-level coding tasks?}

\myparagraph{Motivation} The research question addressing the effectiveness
of the \sysname framework in automating repository-level coding tasks
is of paramount importance in the context of modern software engineering.
Several key motivations drive the significance of this inquiry:
\begin{itemize}
\item \textbf{Complexity of Repository-Level Tasks}: Software engineering
activities, such as package migration and temporal code edits, often
transcend the scope of local code changes. Repository-level coding
tasks involve pervasive modifications across the entire codebase.
This level of complexity necessitates novel approaches to ensure efficiency
and correctness.
\item \textbf{Real-world Relevance}: In practice, software repositories
frequently encounter the need for large-scale changes. For instance,
package migration involves updating dependencies across multiple files
and dependencies, while temporal code edits require tracking and managing
evolving codebase. These tasks are not only time-consuming but
also error-prone when done manually.
\item \textbf{Evaluation against Baselines}: The assessment of \sysname
against baseline methods is crucial. Baseline methods, such as
\textquotedbl Oracle-Guided Repair\textquotedbl{}, are common
in software development but may lack efficiency when dealing with
repository-level tasks. Evaluating \sysname against baselines provides
a benchmark for measuring its effectiveness and highlights areas where
it excels. We also study how our system behaves when using a different
edit model by evaluating a combination of Coeditor and \sysname on the
Temporal Edits Task.
\item \textbf{Large-scale Repositories}: The research considers not only
isolated coding problems but evaluates \sysname's performance on large
internal and external repositories. This broad scope ensures that
the framework's effectiveness is tested under diverse and challenging
real-world scenarios.
\end{itemize}

\myparagraph{Experimental Setup}

To study how well \sysname is able to localize and make the required changes for repository-level tasks, we evaluate it in the context of the tasks described in~\ref{sec:datasets}.
For both the C\# Migration and Temporal Edits tasks, we start with the repository in its \textit{Source} state without any edits having been made.
We apply the seed edits on top of the \textit{Source} state at which point \sysname (or the baseline being evaluated) takes over to complete the the edit across the repository.
In the case of C\# Migrations, the seed edits themselves are performed using the LLM with a suitable prompt, while for Temporal Edits, we the seed edits for each repository task are stored a-priory and applied as patches.
\sysname performs incremental dependency analysis on the repository after the seed edit, identifies code that may be impacted by it, plans which edit to be made next, queries the LLM with a suitable prompt and merges the result back into the repository.
\sysname does this iteratively until it finds that there are no more sites to be edited.

At times, multiple iterations become necessary due to the inherent variability in Large Language Model (LLM) responses. We initiate the first iteration, known as "Iter 1," to kickstart the code editing process using LLMs. However, occasional inaccuracies in LLM responses may introduce erroneous code changes and subsequent build errors. To address these challenges, the second iteration, referred to as "Iter 2," becomes important. During this phase, \sysname actively identifies and acknowledges any build errors from the previous iteration, and re-engagement with the LLM to obtain more precise responses for correcting the initial errors.

Alongside \sysname, we also evaluate a series of baselines on the same repositories.
For C\# migration we evaluate Build-Repair and for Temporal Edits we evaluate Pyright-Repair, the setup for both of which is presented in ~\ref{sec:oracles_baselines}.
Pyright-Strict-Repair is a variant in which we use the Pyright tool with strict mode enabled.
In all the Repair baselines, we provide the same context (temporal and spatial) as in \sysname, the only difference being that the localisation of sites to edit is using the oracle.
We also evaluate using Coeditor instead of gpt-4-32k as the edit model as described in~\ref{sec:oracles_baselines}.
In all Coeditor baselines, contextualisation is performed as in \cite{wei2023coeditor}, with the localisation of next edit site being done through either \sysname or using the oracle.

We evaluate all of these approaches for how well they localise the site to be edited using the \textit{Matched}, \textit{Missed} and \textit{Spurious} Blocks metrics and the correctness of the overall modification using \textit{Levenshtein Distance} and \textit{Diff BLEU} as described in~\ref{sec:metrics}.
We also determine whether the state of the repository after the approach is finished executing passes the validity check -- that is whether it satisfies the oracle and makes the correct edit according to the ground truth.

\begin{table*}[t]
\centering
\resizebox{\textwidth}{!}{
\begin{tabular}{|>{\centering}p{1.5cm}|c|>{\centering}p{1.2cm}|>{\centering}p{1.2cm}|>{\centering}p{1.2cm}|>{\centering}p{1cm}|>{\centering}p{1.7cm}|>{\centering}p{1.3cm}|}
\hline 
\textbf{Dataset} & \textbf{Approach} & \textbf{Matched Blocks} & \textbf{Missed Blocks} & \textbf{Spurious Blocks} & \textbf{Diff BLEU} & \textbf{Levenshtein Distance} & \textbf{Validity Check}\tabularnewline
\hline 
\hline 
\rowcolor{Dandelion}
\multicolumn{8}{|c|}{C\# Migration Task on Internal (Proprietery) Repositories}\tabularnewline
\hline 
\multirow{3}{1.5cm}{Int-1 (Logging)} & \sysname (Iter 1) & \textbf{151} & \textbf{0} & 0 & 0.99 & 60 & \textcolor{red}{\ding{55}} (4)\tabularnewline
\cdashline{2-8}[1pt/1pt]
 & \sysname (Iter 2) & \textbf{151} & \textbf{0} & 0 & 1.00 & 0 & \textcolor{darkgreen}{\ding{51}} (0)\tabularnewline
\cdashline{2-8}
 & Build-Repair & 82 & 69 & 13 & 0.81 & 6465 & \textcolor{red}{\ding{55}} (46)\tabularnewline
\hline
\multirow{3}{1.5cm}{Int-2 (Logging)} & \sysname (Iter 1) & \textbf{438} & \textbf{0} & 0 & 0.99 & 90 & \textcolor{red}{\ding{55}} (6)\tabularnewline
\cdashline{2-8}[1pt/1pt]
 & \sysname (Iter 2) & \textbf{438} & \textbf{0} & 0 & 1.00 & 0 & \textcolor{darkgreen}{\ding{51}} (0)\tabularnewline
\cdashline{2-8}
 & Build-Repair & 337 & 101 & 25 & 0.66 & 7496 & \textcolor{red}{\ding{55}} (68)\tabularnewline
\hline 
\rowcolor{Dandelion}
\multicolumn{8}{|c|}{C\# Migration Task on External (Public) Repositories}\tabularnewline
\hline 
\multirow{2}{1.5cm}{Ext-1 (NUnit-XUnit)} & \sysname (Iter 1) & \textbf{58} & \textbf{0} & 0 & 1.00 & 0 & \textcolor{darkgreen}{\ding{51}} (0) \tabularnewline
\cdashline{2-8}
 & Build-Repair & 52 & 6 & 1 & 0.94 & 530 & \textcolor{red}{\ding{55}} (8)\tabularnewline
 &  &  &  &  &  &  & \tabularnewline
\hline
% \multirow{3}{1.5cm}{Ext-2 (Crypto Library)} & \sysname (Iter 1) &  &  &  &  &  & \tabularnewline
% \cdashline{2-8}[1pt/1pt] 
%  & \sysname (Iter 2) &  &  &  &  &  & \tabularnewline
% \cdashline{2-8}
%  & Build-Repair &  &  &  &  &  & \tabularnewline
% \hline 
% \multirow{3}{1.5cm}{Ext-3 (ASP.NET)} & \sysname (Iter 1) &  &  &  &  &  & \tabularnewline
% \cdashline{2-8}[1pt/1pt] 
%  & \sysname (Iter 2) &  &  &  &  &  & \tabularnewline
% \cdashline{2-8}
%  & Build-Repair &  &  &  &  &  & \tabularnewline
% \hline 
\rowcolor{Dandelion}
\multicolumn{8}{|c|}{Python Temporal Edit Task on External (Public) Repositories}\tabularnewline
\hline 
\multirow{3}{1.5cm}{ T-1 } & \sysname (Iter 1) & \textbf{8} & \textbf{2} & 0 & \textbf{0.90} & \textbf{1044} & \textcolor{red}{\ding{55}}\tabularnewline
\cdashline{2-8} 
 & Pyright-Repair & 5 & 5 & 0 & 0.76 & 1090 & \textcolor{red}{\ding{55}} \tabularnewline
\cdashline{2-8} 
 & Pyright-Strict-Repair & \textbf{8} & \textbf{2} & 0 & \textbf{0.90} & \textbf{1045} & \textcolor{red}{\ding{55}}\tabularnewline
 \cdashline{2-8} 
 & Coeditor-CodePlan & \textbf{8} & \textbf{2} & 0 & \textbf{0.90} & 1160 & \textcolor{red}{\ding{55}} \tabularnewline
\cdashline{2-8} 
 & Coeditor-Pyright-Repair & 5 & 5 & 0 & 0.66 & 1205 & \textcolor{red}{\ding{55}}\tabularnewline
 \cdashline{2-8} 
 & Coeditor-Pyright-Strict-Repair & 5 & 5 & 0 & 0.65 & 1139 & \textcolor{red}{\ding{55}}\tabularnewline
\hline 
\multirow{3}{1.5cm}{ T-2 } & \sysname (Iter 1) & \textbf{4} & \textbf{0} & 0 & \textbf{0.86} & \textbf{248} & \textcolor{darkgreen}{\ding{51}}\tabularnewline
\cdashline{2-8} 
 & Pyright-Repair & 1 & 3 & 0 & 0.00 & 344 & \textcolor{red}{\ding{55}}\tabularnewline
\cdashline{2-8}
 & Pyright-Strict-Repair & 1 & 3 & 0 & 0.00 & 344 & \textcolor{red}{\ding{55}}\tabularnewline
 \cdashline{2-8} 
 & Coeditor-CodePlan (Iter 1) & 3 & 1 & 0 & 0.82 & 254 & \textcolor{red}{\ding{55}}\tabularnewline
\cdashline{2-8} 
 & Coeditor-Pyright-Repair & 1 & 3 & 0 & 0.00 & 344 & \textcolor{red}{\ding{55}}\tabularnewline
 \cdashline{2-8} 
 & Coeditor-Pyright-Strict-Repair & 1 & 3 & 0 & 0.00 & 344 & \textcolor{red}{\ding{55}}\tabularnewline
\hline 
\multirow{3}{1.5cm}{ T-3 } & \sysname (Iter 1) & \textbf{11} & \textbf{0} & 0 & \textbf{0.92} & \textbf{358} & \textcolor{darkgreen}{\ding{51}}\tabularnewline
\cdashline{2-8} 
 & Pyright-Repair & 1 & 10 & 0 & 0.00 & 798 & \textcolor{red}{\ding{55}}\tabularnewline
\cdashline{2-8} 
 & Pyright-Strict-Repair & 1 & 10 & 0 & 0.00 & 798 & \textcolor{red}{\ding{55}}\tabularnewline
 \cdashline{2-8} 
 & Coeditor-CodePlan (Iter 1) & 10 & 1 & 0 & 0.78 & 717 & \textcolor{red}{\ding{55}}\tabularnewline
\cdashline{2-8} 
 & Coeditor-Pyright-Repair & 1 & 10 & 0 & 0.00 & 798 & \textcolor{red}{\ding{55}}\tabularnewline
 \cdashline{2-8} 
 & Coeditor-Pyright-Strict-Repair & 1 & 10 & 0 & 0.00 & 798 & \textcolor{red}{\ding{55}}\tabularnewline
\hline 
\end{tabular}}
\caption{Comparison of \sysname's repository edit metrics with Build-Repair baseline. Higher values of Matched Blocks, Diff BLEU, and lower values of Missed Blocks, Spurious Blocks, Levenshtein Distances are better.}
\label{tab:migration_results}
\end{table*}

\myparagraph{Results Discussion}
The experimental results in Table \ref{tab:migration_results} demonstrate the effectiveness of the \sysname
framework in automating repository-level coding tasks.

In the context of the C\# Migration Task on Internal (Proprietary) Repositories, the results table provides a comprehensive view of the performance of two approaches: \sysname and Build-Repair. Notably, \sysname demonstrates superior capabilities in several key aspects. It excels in "Matched Blocks", achieving perfect results with 151 matched blocks for both "Int-1 (Logging)" and 438 matched blocks for "Int-2 (Logging)" datasets. This indicates \sysname's exceptional precision in accurately identifying and addressing intended code changes. Moreover, \sysname impressively exhibits zero "Missed Blocks" ensuring that no crucial code modifications are overlooked, thus minimizing the risk of functional issues. Equally noteworthy is the absence of "Spurious Blocks" introduced by \sysname, signifying its ability to maintain codebase cleanliness and integrity. In contrast, Build-Repair falls behind with 82 matched blocks for "Int-1 (Logging)" and 437 for "Int-2 (Logging)", while also missing a substantial number of blocks (69 and 101, respectively). Additionally, Build-Repair introduces 13 spurious blocks for "Int-1 (Logging)" and 25 for "Int-2 (Logging)", which can increase code complexity and error potential. These findings underscore \sysname's superiority over Build-Repair, not only in terms of matched blocks but also in its ability to avoid missed and spurious code changes, ultimately offering a more precise and reliable solution for the C\# Migration Task on internal repositories.

\myparagraph{\sysname vs Build-Repair} The comparative analysis reveals why Build-Repair falls behind \sysname. One crucial factor contributing to its performance gap is its reliance on "build error location" as an indicator for code correction. Build errors often pinpoint the location where an error is detected, but they may not necessarily coincide with the actual required fix location. For instance, an error may manifest in a derived class's overridden function signature mismatch, but the fix is necessitated in the base class's virtual function signature, causing Build-Repair to misinterpret the correction site. Additionally, the build process in the context of compiler optimizations may obscure subsequent errors, only displaying selected errors at a given time. This can lead to an incorrect identification of the correction location and hinder the proper propagation of changes, further exacerbating the performance gap between \sysname and Build-Repair. These limitations underscore the challenges faced by Build-Repair in accurately pinpointing and addressing code modifications in complex migration tasks.

\myparagraph{Multiple Iterations} We can see the necessity for multiple iterations to handle the inherent variability in Large Language Model (LLM) responses as illustrated with the 4 build errors after "Iter 1" of \sysname in the "Int-1" dataset. To rectify this, "Iter 2" plays a vital role. In this phase, \sysname identifies and acknowledges the build errors from the previous iteration and re-engages with the LLM to obtain more accurate responses for correcting the initial errors. It is noteworthy that there are no changes to the block metrics between the two iterations since \sysname correctly identifies the blocks requiring correction and engages the LLM for modification. However, the LLM corrections in "Iter 1" were erroneous, leading to a lower Levenshtein distance metric. This iterative refinement process significantly contributes in mitigating the impact of occasional inaccuracies in LLM outputs during the code editing phases.

\myparagraph{Performance on Ext-1} In the comparison between CodePlan and the Build-Repair baseline method on the "Ext-1" dataset, we observe a significant difference in their performance. CodePlan successfully identified and updated 58 code blocks, achieving a perfect DiffBLEU score of 1.00, indicating that it made changes identical to the target repository. In contrast, Build-Repair, the baseline method, fell short by missing six blocks and generating eight build errors. This discrepancy highlights a critical limitation of Build-Repair—the lack of comprehensive change may-impact analysis capabilities. Specifically, Build-Repair failed to update constructor blocks required for initializing the newly added \code{ITestOutputHelper \_output} class member. This omission went unnoticed because the absence of initialization didn't trigger build errors, causing a cascading effect on the callers of this constructor. In contrast, CodePlan successfully handled the initialization of the newly added \code{ITestOutputHelper \_output} class member. This achievement can be attributed to its robust change-may impact analysis, which accurately identified the necessary modifications to the constructor block upon the addition of a new field. As a result, CodePlan seamlessly updated the constructor and all its callers, preventing any missed blocks or build errors. This finding underscores the importance of CodePlan's advanced planning abilities, which ensure a more thorough and accurate approach to repository-level code edits. Further details and qualitative analysis are provided in RQ3 (Figure \ref{fig:codeplan_change_propagation}) of our study.

\myparagraph{\sysname vs Pyright-Repair} In context of the Temporal Edits task, we can see that \sysname is able to successfully identify all the derived edit location in 2 repos (T-2, T-3) and is almost successuful in the third (T-1). 
This is in contrast to the baseline Pyright-Repair approach which fails to identify any of the derived edits in two repos (T-2, T-3).
We find that using Pyright on a strict checking mode yields improved results, but only on one repo (T-1) where it peforms just as well as \sysname. 
Overall we observe that the Pyright-Repair baseline falls short in identifying location to edit.
This is also reflected in the DiffBLEU score which is consistently higher for \sysname and Levenshtein Distance (L.D.) which is consistently lower.
Note that since the LLM does not necessarily make the exact edit present in the ground truth, both DiffBLEU and L.D. may not have perfect 1.0 and 0 values respectively.

\begin{figure}
\begin{minipage}{0.47\linewidth}
% \begin{codebox}{Task Description}
% \raggedright
\begin{lstlisting}[style=diff,frame=single]
def load_mbd_ckpt(
    file_or_url_or_id: Union[Path, str],
+   filename: tp.Optional[str] = None,
    cache_dir: tp.Optional[str] = None
):
    return _get_state_dict(
        file_or_url_or_id,
-       filename="all_in_one.pt",
+       filename=filename,
        cache_dir=cache_dir
)
\end{lstlisting}  
\vspace*{-7pt}
\caption{Seed edit for T-2.}
\label{fig:t2_seed}
\end{minipage}
\hfill
\begin{minipage}{0.47\linewidth}
\centering
\begin{lstlisting}[style=diff,frame=single]
def send_request(data):
    api_key = data.pop("api_key")
    api_type = data.pop("api_type")
+   api_endpoint = data.pop("api_endpoint")
    ...
\end{lstlisting}
\vspace*{-7pt}
\caption{Seed edit for T-3.}
\label{fig:t3_seed}
\end{minipage}
\end{figure}

For both T-2 and T-3 we see that the Pyright-Repair baseline is unable to identify any derived edits.
In both of these repos Pyright does not flag any errors in the codebase once the seed-edit has been made.
In the case of T-2, the seed edit involves adding an argument to a method as shown in Figure~\ref{fig:t2_seed}
However this new parameter is also assigned a default value.
Due to the presence of a default value for this argument, Pyright does not flag any of the call sites of this method as errors.
However in the ground truth, the developer does follow up and edit the call sites.
\sysname's change may-impact analysis identifies these call-sites as "might requiring" edit and so we pass them to the LLM to edit. 

In the case of T-3, the seed edit involves modifying the body of the function as describe in Figure~\ref{fig:t2_seed}.
After the seed edit, the method now expects the dictionary being passed as argument to contain a new key {\tt "api\_endpoint"}.
Pyright will not flag any errors in the repo at this stage however it is clear that this may require modification to the callers of {\tt send\_request}.
\sysname does detect this and is able to identify and make edits at 10 derived sites present in the ground truth.

In both of these scenarios, we do not know a-priory whether the call-sites need editing.
Thus \sysname leaves this decision upto the LLM to decide given the context.
In contrast, the Oracle-Guided baseline does not even detect that a change maybe needed.
This can be attributed to the fact that Oracles such as Pyright, or Build aim to detect errors and hence will only flag code that violates certain rules.
This is not aligned with the task of propagating changes across a repo, as there may be many cases where a change may need to be propagated but the repo with the change applied does violate any of the oracle's rules.

\myparagraph{Coeditor Evaluation} When comparing \sysname with Coeditor-CodePlan, we can see that it performs on par on T-1 but lags slightly behind on T-2 and T-3 missing one edit site in each. 
While both approaches are using the same analysis and planning, the local edits made in \sysname are more in-line with the ground-truth compared to those by Coeditor, reflected in the the lower DiffBLEU scores and higher L.D. exhibited by Coeditor-CodePlan in T-2 and T-3.
This can be due to the difference in context-understanding abilities between gpt-4-32k and Coeditor.
For example, opting to instantiate an argument to a method instead of adding a parameter to the caller can mean missing out of edits resulting from the signature change to the caller.
Being a significantly more powerful model, gpt-4-32k is better at understanding the context of the temporal edits, hence the edits it makes are more aligned with the ground truth as compared to Coeditor.
This is also observed when comparing the Pyright-Strict-Repair and Coeditor-Pyright-Strict-Repair on T-2, where incorrect local edits by Coeditor lead to missing of edit sites and worse DiffBLEU and L.D.

In conclusion, the experimental results substantiate that \sysname's planning-based approach is highly effective in automating repository-level coding tasks, offering superior matching, completeness, and precision compared to traditional baseline methods. Its ability to handle complex coding tasks within large-scale repositories signifies a significant advancement in automating software engineering activities.

\subsection{RQ2: How important are the temporal and spatial contexts for \sysname's performance?}

\myparagraph{Motivation}: 
The motivation behind RQ2 lies in recognizing that Large Language Models (LLMs) require both temporal and spatial contexts to provide accurate and contextually relevant code update suggestions. Temporal context is vital for understanding the sequence and timing of code changes, allowing LLMs to make suggestions that align with the code's evolution and maintain consistency. Without this temporal awareness, LLMs might offer solutions that conflict with earlier or subsequent modifications, leading to code errors. Spatial context, on the other hand, enables LLMs to comprehend the intricate relationships and dependencies between different parts of the codebase. This understanding is crucial for pinpointing where updates are needed and ensuring that they are applied in a way that preserves code functionality. Therefore, investigating the importance of these contexts in \sysname's planning is fundamental to harnessing the full potential of LLMs in automating repository-level coding tasks effectively.

\myparagraph{Experimental Setup}: To assess the importance of temporal and spatial contexts in \sysname's planning (RQ2), a specific experimental setup is employed. Recall that temporal contexts are tracked by \sysname through the maintenance of a list of previous related changes, and these contexts are subsequently incorporated into the Large Language Model (LLM) prompt.

To measure the significance of these contexts, a controlled experiment is carried out. In this experiment, the tracking of temporal changes is intentionally halted, and no temporal or spatial contexts are provided to the LLM during code modification tasks. This enables a detailed examination of how LLM responses are impacted when these essential contexts are omitted. The experiment also encompasses a comprehensive assessment of various metrics, including code consistency (block metrics) and code correctness (Diff BLEU and Levenshtein distance). By comparing the outcomes between \sysname's setup with temporal and spatial contexts and the experimental setup without them, the research aims to precisely quantify the importance of these contexts in \sysname's planning process and their influence on the quality of automated code modifications.

\myparagraph{Results Discussion}: 

The results regarding the importance of temporal and spatial contexts for \sysname's planning (RQ2) reveal critical insights. As observed in Table 2, when temporal contexts are not considered, there is a noticeable increase in missed blocks during the code modification process. This increase is attributed to the Large Language Model (LLM) not making necessary changes to certain code blocks due to its inability to comprehend the need for those modifications in the absence of temporal context.

\begin{table*}[t]
\centering
\begin{tabular}{|>{\centering}p{1.5cm}|c|>{\centering}p{1.2cm}|>{\centering}p{1.2cm}|>{\centering}p{1.2cm}|>{\centering}p{1.2cm}|>{\centering}p{1.7cm}|>{\centering}p{1.2cm}|}
\hline 
\textbf{Dataset} & \textbf{Approach} & \textbf{Matched Blocks} & \textbf{Missed Blocks} & \textbf{Spurious Blocks} & \textbf{Diff BLEU} & \textbf{Levenshtein Distance} & \textbf{Oracle Verdict}\tabularnewline
\hline 
\hline 
\rowcolor{Dandelion}
\multicolumn{8}{|c|}{C\# Migration Task on Int-1 with Temporal and Spatial Contexts}\tabularnewline
\hline 
\multirow{2}{1.5cm}{Int-1} & \multirow{2}{*}{\begin{tabular}[c]{@{}c@{}}\sysname \\ (Iter 1 + Iter 2)\end{tabular}} & \multirow{2}{*}{151} & \multirow{2}{*}{0} & \multirow{2}{*}{0} & \multirow{2}{*}{1.00} & \multirow{2}{*}{0} & \multirow{2}{*}{\textcolor{darkgreen}{\ding{51}} (0)}\tabularnewline
%\cline{2-8} \cline{3-8} \cline{4-8} \cline{5-8} \cline{6-8} \cline{7-8} \cline{8-8} 
 &  &  &  &  &  &  & \tabularnewline
\hline 
\rowcolor{Dandelion}
\multicolumn{8}{|c|}{C\# Migration Task on Int-1 \textbf{without} Temporal and Spatial Contexts}\tabularnewline
\hline 
\multirow{3}{1.5cm}{Int-1} & \sysname (Iter 1) & 112 & 39 & 4 & 0.73 & 3674 & \textcolor{red}{\ding{55}} (34) \tabularnewline
\cline{2-8} \cline{3-8} \cline{4-8} \cline{5-8} \cline{6-8} \cline{7-8} \cline{8-8} 
 & \sysname (Iter 2) & 121 & 30 & 52 & 0.53 & 4522 & \textcolor{red}{\ding{55}} (68) \tabularnewline
\cline{2-8} \cline{3-8} \cline{4-8} \cline{5-8} \cline{6-8} \cline{7-8} \cline{8-8} 
 & \sysname (Iter 3) & 121 & 30 & 54 & 0.51 & 4524 & \textcolor{red}{\ding{55}} (69)\tabularnewline
\hline 
\end{tabular}
\caption{Ablation study with and without temporal/spatial blocks. Without temporal/spatial contexts, \sysname is unable to make correct edits.}
\label{tab:ablation_results}
\end{table*}

\begin{figure*}[t]
    \centering
    \scalebox{1}{
    \includegraphics[width=1.0\textwidth]{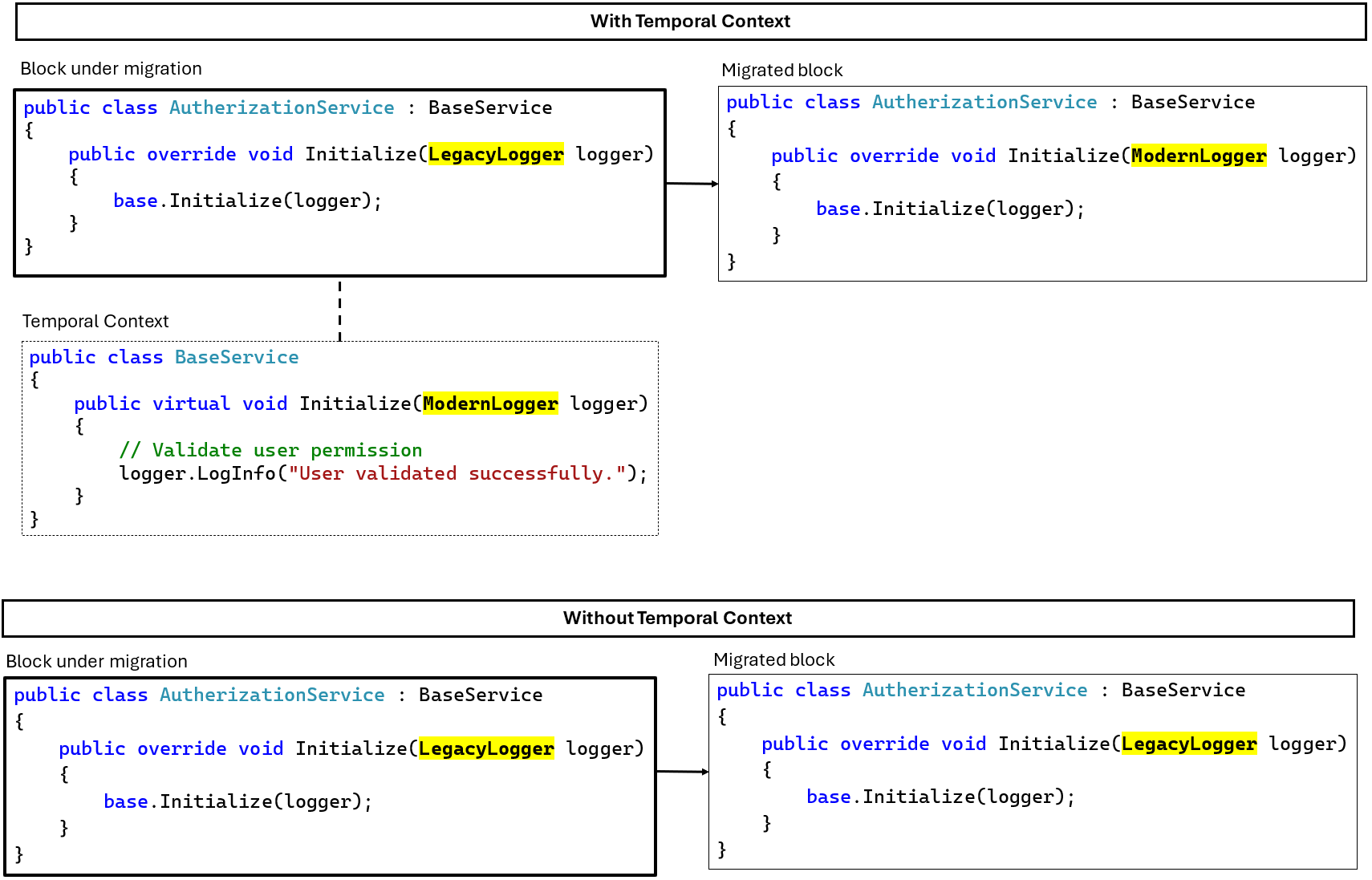}}
    \vspace*{-10pt}
    \caption{Illustration of importance of temporal context. Failure to update LogacyLogger to ModernLogger in Initialize() method is the results of missing missing temporal context.}
    \label{fig:temporal_1}
\end{figure*}

An illustrative example in Figure \ref{fig:temporal_1} exemplifies this issue. In this scenario, a correction is required in the base class's virtual method based on changes to the overridden method's signature in the derived class. However, the LLM, lacking temporal context, does not possess information about the derived class's method, leading it to believe that no changes are necessary to the base class method. This highlights the critical role that temporal context plays in understanding code dependencies and ensuring accurate updates.

\begin{figure*}[t]
    \centering
    \scalebox{1}{
    \includegraphics[width=1.0\textwidth]{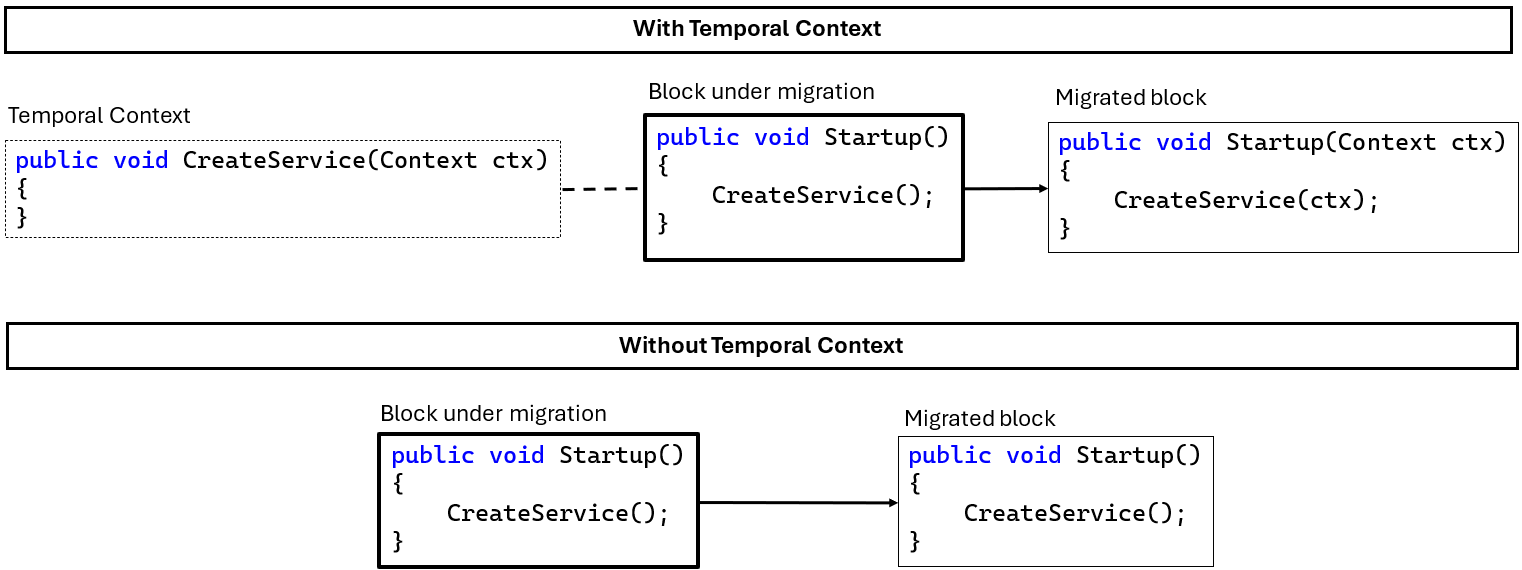}}
    \vspace*{-10pt}
    \caption{Illustration of importance of temporal context. Failed update to Startup() method is the results of missing missing temporal context.}
    \label{fig:temporal_2}
\end{figure*}

Furthermore, Figure \ref{fig:temporal_2} provides another instance where the absence of temporal context impacts the code modification process. In this case, a "Context" parameter needs to be added to the "CreateService()" call within the "Startup()" method. However, since the LLM lacks temporal context, it is unaware of the signature change to "CreateService()" and, consequently, fails to recognize the need for updates to all the callers. This omission results in numerous missed updates throughout the codebase.

It's crucial to highlight another significant observation: the increase in the count of spurious blocks when spatial context is insufficient. This phenomenon occurs because, in the absence of adequate spatial context, the Large Language Model (LLM) may incorrectly perceive missing code elements and attempt to create them, leading to the generation of spurious code blocks.

\begin{figure*}[t]
    \centering
    \scalebox{1}{
    \includegraphics[width=1.0\textwidth]{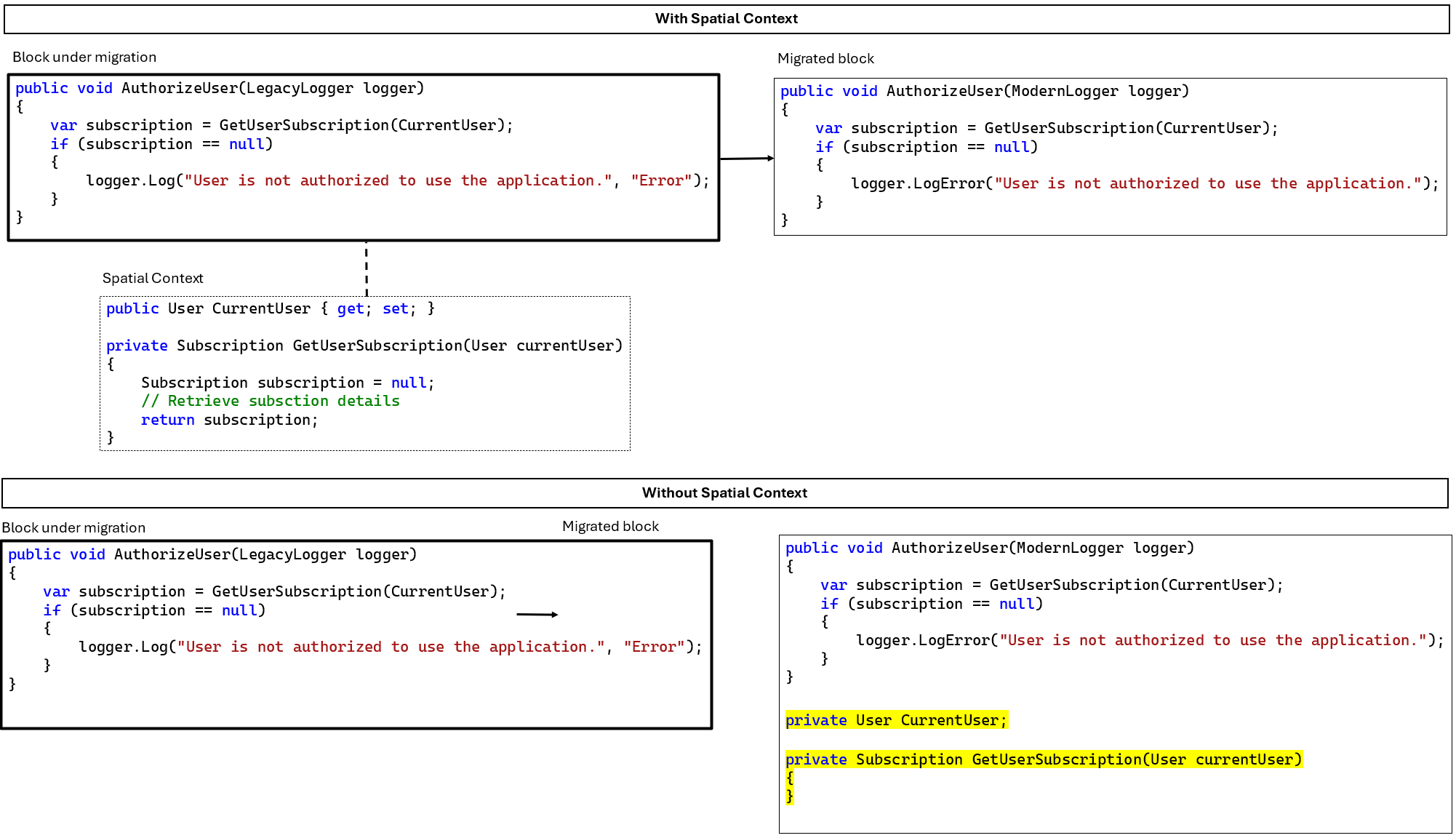}}
    \vspace*{-10pt}
    \caption{Illustration of importance of spatial context. Spurious blocks, highlighted in yellow are the results of missing missing spatial context.}
    \label{fig:spatial_1}
\end{figure*}

An illustrative example in Figure \ref{fig:spatial_1} demonstrates this issue. In this scenario, the task is to modify the "AuthorizeUser()" method by migrating the logging calls from an old logging framework to a new one. However, due to the lack of spatial context that would specify the existence of the "GetUserSubscription()" method and the "CurrentUser" property, the LLM attempts to create these elements as well. Consequently, not only is the logging migration addressed, but the LLM also introduces unnecessary code blocks, such as the creation of the "GetUserSubscription()" method and the addition of "CurrentUser" as a class-level object.

This observation underscores the critical role of spatial context in guiding the LLM's understanding of code structure and relationships. Providing comprehensive spatial context can help prevent the generation of superfluous code blocks and ensure that code modifications are precise and aligned with the intended changes.

In summary, the experimental results emphasize the essential nature of temporal and spatial contexts in \sysname's planning. The increase in missed and spurious updates due to the absence of temporal and spatial contexts underscores the significance of providing the LLM with a comprehensive understanding of code evolution and dependencies through these contexts to ensure accurate and effective code modifications.

\subsection{RQ3: What are the key differentiators that allow \sysname to outperform baselines in solving complex coding tasks?}

\myparagraph{Motivation}: RQ3 seeks to uncover the key differentiators that empower \sysname to excel in tackling intricate coding tasks compared to baseline approaches. This research question is motivated by the need to identify and understand the factors that contribute to \sysname's superior performance. Given the increasing complexity of coding tasks, especially those at the repository level, it becomes crucial to pinpoint the specific aspects that set \sysname apart from traditional methods. By discerning these differentiators, the research aims to shed light on the strengths and advantages of \sysname, offering valuable insights into how it effectively addresses the challenges posed by complex coding tasks.

\myparagraph{Experimental Setup}: The primary focus here is on qualitative analysis. After conducting code modifications with both \sysname and the baseline methods, the results are meticulously examined through manual inspection. This entails a detailed examination of the changes made by each approach, with the aim of gaining qualitative insights into the decision-making processes and the nuances of code modifications. By manually eyeballing and comparing the alterations, the research seeks to uncover subtle yet crucial distinctions that illuminate the strengths and underlying mechanisms that give \sysname its edge in addressing the challenges posed by intricate coding tasks. This qualitative approach provides a comprehensive understanding of how \sysname excels in this context and what sets it apart from conventional methods.

\myparagraph{Results Discussion}: 

\myparagraph{\sysname's Strategic Planning and Context Awareness:}

\begin{figure*}[t]
    \centering
    \scalebox{1}{
    \includegraphics[width=1.0\textwidth]{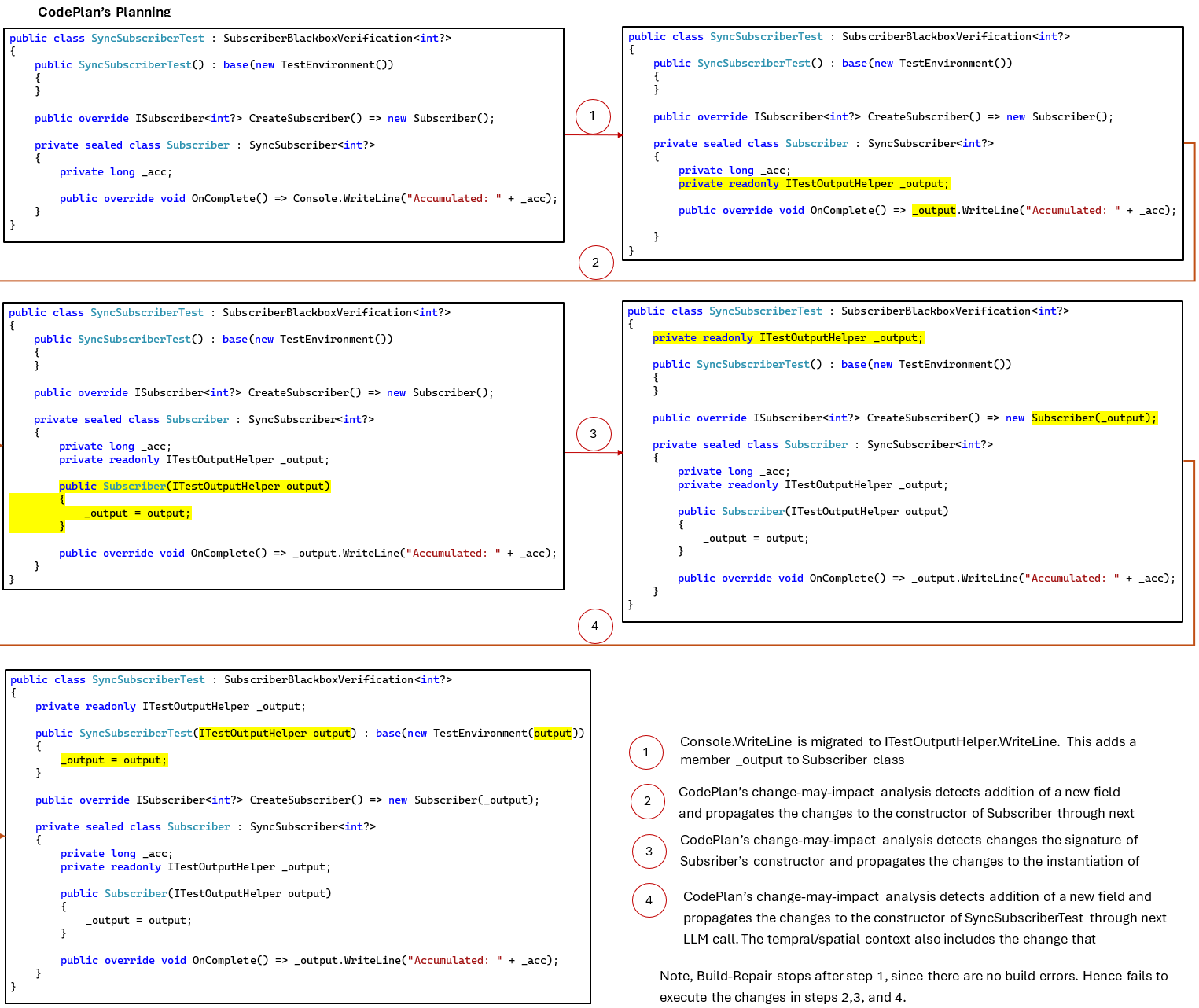}}
    \vspace*{-10pt}
    \caption{Illustration of the \sysname's plan execution.}
    \label{fig:codeplan_change_propagation}
\end{figure*}

\sysname's exceptional performance in handling complex coding tasks can be attributed to its powerful features, notably its incremental analysis and change-may-impact analysis. These capabilities set it apart from baseline methods like Build-Repair, which primarily focus on maintaining syntactic correctness while overlooking critical contextual details. To illustrate this, let's delve into an example from repository Ext-1 illustrated in Figure \ref{fig:codeplan_change_propagation}, where \sysname is tasked with migrating the \code{Console.WriteLine} method to \code{ITestOutputHelper.WriteLine}. This migration involves a series of changes 1 to 4 as described in the Figure \ref{fig:codeplan_change_propagation}. These cascading changes start from introducing \code{ITestOutputHelper \_output} as a class-level member, accomplished via LLM updates.

\sysname's change-may-impact analysis proves invaluable in this scenario. It recognizes that the addition of a new field necessitates modifications to the constructor to ensure proper initialization. As a result, \sysname schedules the necessary constructor modification. Consequently, the constructor \code{Subscriber(...)} is correctly updated to accept \code{ITestOutputHelper} as a parameter and initialize the class member \code{\_output}. This in turn results in a series of changes through the repository as explained in steps 1 to 4 in the Figure \ref{fig:codeplan_change_propagation}.

This example demonstrates how \sysname makes methodical and contextually-aware changes to the repository, thanks to its ability to do change impact analysis and incorporate temporal contexts. In contrast, Build-Repair, reliant solely on syntactic correctness, fails to even detect the need for modification in the Subscriber's constructor. Given that all syntactic rules are adhered to, it does not prompt a build error and consequently fails to implement changes in steps 2 to 4, as illustrated in Figure 4. Instead, it solely executes the modification outlined in step 1, resulting in incomplete code updates.

\sysname's distinctive advantage lies in its holistic understanding of code relationships and its meticulous planning, which ensures the integrity and functionality of the codebase are maintained throughout complex coding tasks. This qualitative analysis highlights how \sysname's nuanced approach outperforms baselines in handling intricate coding challenges.

\myparagraph{Incremental Analysis: Maintaining Relationships with Dependency Graph:}

\sysname's exceptional performance in tackling complex coding tasks is attributed to its incremental analysis, which effectively links edits with the underlying dependency graph. Unlike a static snapshot of code, which may result in an incomplete representation of dependencies, our incremental analysis method ensures that relationships within the dependency graph are maintained until the affected blocks are modified.

Consider a scenario where a caller function undergoes a renaming process. Traditional static snapshots would struggle to preserve the caller-callee relationship because, in their view, the caller has already been renamed. However, \sysname's incremental analysis steps in, preserving the caller-callee relation until the caller function itself undergoes an update. This dynamic approach ensures that critical relationships aren't prematurely severed, allowing for more accurate and context-aware code modifications.

Another instance of \sysname's prowess lies in handling modifications to import statements. Suppose an import statement originally reads as \code{import numpy}, and it's modified to \code{import numpy as np}. In a static snapshot, this alteration could result in the loss of the "ImportedBy" relationship. However, \sysname's incremental analysis ensures that such vital relationships are maintained, facilitating precise and comprehensive code updates.

\myparagraph{Incremental Analysis: Enhanced Spatial and Temporal Context Extraction:}

\sysname's remarkable success in complex coding tasks can be attributed to its proficiency in extracting spatial context more accurately, thanks to incremental analysis. Attempting to extract spatial context without the support of incremental analysis often leads to a loss of accuracy and completeness.

Consider a scenario where a method within the codebase constructs an object of a class, let's say "A." However, at some point in the code's history, "A" was renamed to "B." Traditional methods that lack incremental analysis may struggle with this situation. When attempting to extract the class definition, they may encounter a roadblock because, in the current static snapshot, "A" no longer exists.

However, \sysname's incremental analysis comes to the rescue by establishing the crucial link between the historical context and the present state. It accurately extracts the class definition, recognizing that the object is now of class "B" due to the earlier temporal edit (the renaming of "A" to "B"). This holistic approach ensures that spatial context extraction is both precise and comprehensive, allowing \sysname to make informed and context-aware code modifications.

\myparagraph{Change-may-impact analysis propagates subtle behavioral changes.}

One of the key factors differentiating \sysname's superior performance in complex coding tasks is its adeptness at detecting subtle behavioral changes through extensive change-may-impact analysis. While certain code edits, like modifying method signatures, result in obvious breaking changes that can be detected by build tools, others induce more nuanced behavioral shifts without directly breaking the build. These subtle alterations, often overlooked, can significantly affect code correctness and functionality. For instance, a seemingly minor change in a method's return value, from True to False, may invalidate assertions in unit tests.

\sysname excels in identifying such subtle behavioral transformations that may elude oracles such as build or static checking tools. Its thorough change-may-impact analysis delves beyond surface-level modifications, proactively recognizing these inconspicuous shifts. This capability sets \sysname apart from baseline methods, which primarily focus on changes related to build success. Consequently, \sysname emerges as a powerful solution for addressing complex coding tasks, ensuring that even the most subtle alterations are meticulously considered, ultimately enhancing code quality.

\myparagraph{Change may-impact analysis maintains cause-effect relationship.}
One of \sysname's differentiators lies in its proficiency in preserving the cause-effect relationship when handling complex coding tasks. Traditional build tools are effective at pinpointing breaking changes but often fall short in identifying the underlying causes and their corresponding effects. For instance, if a method signature is altered within an overridden method, a typical build tool would flag the issue at the overridden method's location, where the error is observed. However, this approach fails to recognize the underlying cause—the change in the method signature, which should ideally lead to an update in the corresponding virtual method in the base class.

In contrast, \sysname's change-may-impact analysis excels in maintaining the causal link between code modifications. When a breaking change is introduced, \sysname not only identifies the error but also traces it back to the root cause, establishing the need for subsequent changes. In the aforementioned example, \sysname recognizes that the change in the overridden method's signature necessitates an update to the corresponding virtual method in the base class. This meticulous preservation of cause and effect sets \sysname apart from baseline methods, which often treat issues in isolation without considering the broader context.

\myparagraph{Incremental static analysis is lightweight and easy to deploy.}
One of the key distinguishing features that enables \sysname to outperform baselines in tackling complex coding tasks is its utilization of lightweight and readily deployable incremental static analysis. In real-world software development, build systems often prove to be complex and resource-intensive. Traditional build processes can be quite expensive in terms of computational resources and time, especially when dealing with large and intricate codebases.

In contrast, \sysname's incremental static analysis offers a more efficient and practical alternative. By focusing on analyzing only the code portions that have been modified, \sysname significantly reduces the computational overhead associated with build processes. This approach not only accelerates the repository-wide editing but also minimizes the resources required for each editing task. 

\smallskip

In summary, \sysname's exceptional performance in complex coding tasks can be attributed to several key differentiators. Firstly, its strategic planning, coupled with its deep understanding of temporal and spatial contexts, enables \sysname to make methodical and context-aware code modifications. Additionally, the system's ability to maintain relationships within the dependency graph through incremental analysis ensures the preservation of code correctness and functionality. \sysname's enhanced spatial and temporal context extraction further aids in accurately updating code elements. Notably, the change-may-impact analysis not only identifies subtle behavioral changes but also maintains a clear cause-effect relationship between code modifications. Finally, \sysname's lightweight and easily deployable incremental static analysis offer practical advantages, reducing resource overhead in complex software projects. These combined differentiators empower \sysname to excel in solving intricate coding challenges.

\section{Limitations and Threats to Validity}

While \sysname demonstrates significant capabilities, there are certain limitations and potential threats to the validity of its results. One crucial factor for \sysname's success lies in the quality of the dependency analysis. In statically typed languages like C\# and Java, rich code dependency information can be extracted effectively. However, in dynamically typed languages such as Python (without type hints) or JavaScript, establishing semantically rich relationships between code blocks can be more challenging due to the dynamic nature of the code.

Our current implementation of \sysname primarily handles relations between code blocks through static analysis. In real-world software systems, numerous dynamic dependencies exist, including data flow dependencies, complex dynamic dispatching (via virtual functions and dynamic castings), algorithmic dependencies (e.g., when input lists are expected to be sorted), and various execution dependencies (such as multi-threading and distributed processing). Addressing these dynamic dependencies will be a crucial focus of our future work.

We selected multiple repositories across two challenging tasks (migration and temporal edits) and two languages (C\# and Python) to evaluate the generality of \sysname. These repositories and tasks are representative of real-world usecases. However, given the complexity of setting up each experiment, our evaluation is restricted to a few repositories. More experimentation will be required to probe the strengths and weaknesses of \sysname. 

While \sysname excels in planning and editing for programming language repositories, enterprise software systems often comprise various other artifacts like configuration files, metadata files, project setting files, external dependencies, and more. A comprehensive repository-level editing approach, particularly for tasks like software migrations, necessitates the ability to edit these additional artifacts. This calls for an evolution of dependency graphs to include nodes and relations encompassing all these diverse artifacts, and change-may-impact analysis that spans across them.

\sysname currently relies on Large Language Models (LLMs) to utilize the necessary context information for making code edits. It trusts the LLM's response for change-may-impact analysis. In cases where the LLM response is incorrect or spurious, it may lead to erroneous updates in the repository. While running \sysname in multiple iterations in our experiments helped rectify such issues, there may be instances where this approach may not suffice.

In terms of \sysname's update strategy, it currently focuses on updating one block at a time. While this approach aligns with our current design choices, there may be scenarios where it would be more efficient or necessary to implement multiple simultaneous changes. More sophisticated techniques for handling multiple block updates (within the same file or across different files) and change propagation will be a key area of exploration in our future work.

Although our current methodology employs zero-shot prompting, there exists potential for extension to include few-shot examples, Chain of Thought (CoT), and other techniques. These extensions represent a promising avenue for future research.

\section{Related Work}

\myparagraph{LLMs for Coding Tasks}
A multitude of LLMs~\cite{ahmad2021unified,wang_codet5_2021,wang_gpt-j-6b_2021_gpt_j,brown2020language,austin_program_2021_prog_synth_with_lm,chen_evaluating_2021_Codex,black_gpt-neox-20b_2022_gpt_neox,xu_systematic_2022_polycoder,chowdhery2022palm,fried_incoder_2022,openai2023gpt4,touvron2023llama} have been trained on large-scale corpora of source code and natural language text. These have been used to accomplish a variety of coding tasks. A few examples of their use include program synthesis~\cite{li_competition-level_2022_alphaCode,nijkamp_codegen_2023}, program repair~\cite{xia2023automated,jin2023inferfix,ahmed2023majority}, vulnerability patching~\cite{pearce2022examining}, inferring program invariants~\cite{pei2023can}, test generation~\cite{schafer2023adaptive} and multi-task evaluation~\cite{tian2023chatgpt}. However, these investigations are performed on curated examples that are extracted from their repositories and are meant to be accomplished with independent invocations of the LLM. We consider a different class of tasks posed at the scale of code repositories, where an LLM is called multiple times on different examples which are inter-dependent. We monitor the results of each LLM invocation within the repository-wide context to identify future code change obligations to get the repository to a consistent state, e.g., where the repository is free of build or runtime errors.

\myparagraph{Automated Planning}
Automated planning~\cite{ghallab2004automated,russell2010artificial} is a well-studied topic in AI.
Online planning~\cite{russell2010artificial} is used when the effect of actions is not known and the state-space cannot be enumerated \emph{a priori}.
It requires monitoring the actions and plan extension. In our case, the edit actions are carried out by an LLM whose results cannot be predicted before-hand and the state-space is unbounded. As a consequence, our adaptive planning is an online algorithm where we monitor the actions and extend the plan through static analysis. In orthogonal directions, \cite{jiang2023selfplanning} uses an LLM to derive a plan given a natural language intent before generating code to solve complex coding problems and \cite{zhang2023planning} performs lookahead planning (tree search) to guide token-level decoding of code LMs. Planning in our work is based on analyzing dependency relations and changes to them as an LLM makes changes to a code repository.

\myparagraph{Analysis of Code Changes}
Static analysis is used for ensuring software quality. It is expensive to recompute the analysis results every time the code undergoes changes. The field of incremental program analysis offers techniques to recompute only the analysis results impacted by the change. Specialized algorithms have been developed for dataflow analysis~\cite{ryder1983incremental,arzt2014reviser}, pointer analysis~\cite{yur1999incremental}, symbolic execution~\cite{person2011directed}, bug detection~\cite{mcpeak2013scalable} and type analysis~\cite{busi2019using}. Program differencing~\cite{apiwattanapong2004differencing,lahiri2012symdiff,kim2012identifying} and change impact analysis~\cite{arnold1996introduction,jashki2008towards} determine the differences in two program versions and the effect of a change on the rest of the program. The impact of changes has been studied for regression testing~\cite{ren2004chianti}, analyzing refactorings~\cite{dig2006automated} and assisting in code review~\cite{alves2014refdistiller,ge2017refactoring}.
We analyze the code generated by an LLM and incrementally update the syntactic (e.g., parent-child) and dependency (e.g., caller-callee) relations. We further analyze the likely impact of those changes on related code blocks and create change obligations to be discharged by the LLM.

\myparagraph{Spatial and Temporal Contextualization}
As discussed in the Introduction, LLMs benefit from relevant context derived from other files in the repository and from past edits. We provide both these pieces of information to the LLM by tracking the code changes and dependency relations.

\myparagraph{Learning Edit Patterns}
Many approaches have been developed to learn edit patterns from past edits or commits in the form of rewrite rules~\cite{revisar}, bug fixes~\cite{andersen2010generic,getafix}, type changes~\cite{ketkar2022inferring}, API migrations~\cite{lamothe2020a3,xu2019meditor} and neural representations of edits~\cite{yin2019learning}. Approaches such as \cite{meng2011sydit} and \cite{meng2013lase} synthesize context-aware edit scripts from user-provided examples and apply them in new contexts. Other approaches observe the user actions in an IDE to automate repetitive edits~\cite{miltner2019fly} and temporally-related edit sequences~\cite{overwatch}. We do not aim to learn edit patterns and we do not assume similarities between edits. Our focus is to identify effects of code changes made by an LLM and to guide the LLM towards additional changes that become necessary.

\section{Conclusions and Future Work}

In this paper, we introduced \sysname, a novel framework designed to tackle the challenges of repository-level coding tasks, which involve pervasive code modifications across large and inter-dependent codebases. \sysname leverages incremental dependency analysis, change may-impact analysis, and adaptive planning to orchestrate multi-step edits guided by Large Language Models. We evaluated \sysname on diverse code repositories with varying complexities and sizes, including both internal proprietary repositories and public GitHub repositories in C\# and Python for migration and temporal edit tasks. Our results demonstrated that \sysname outperforms baseline methods, achieving better alignment with the ground truth.
In conclusion, \sysname presents a promising approach to automating complex repository-level coding tasks, offering both productivity benefits and accuracy improvements. Its success in addressing these challenges opens up new possibilities for efficient and reliable software engineering practices.

While \sysname has shown significant promise, there are several avenues for future research and enhancements. First, we aim to expand its applicability to a broader range of programming languages and code artifacts, including configuration files, metadata, and external dependencies, to provide a more holistic solution for repository-level editing.
Additionally, we plan to explore further customization of \sysname's change may-impact analysis. This could involve incorporating task-specific impact analysis rules, either through rule-based methods or more advanced machine learning techniques, to fine-tune its editing decisions for specific coding tasks. Furthermore, we will address the challenge of handling dynamic dependencies, such as data flow dependencies, complex dynamic dispatching (via virtual functions and dynamic castings), algorithmic dependencies (e.g., when input lists are expected to be sorted), and various execution dependencies (such as multi-threading and distributed processing), to make \sysname even more versatile in addressing a wider range of software engineering tasks.

\bibliographystyle{ACM-Reference-Format}
\bibliography{references}

\end{document}